\documentclass[draft]{agujournal2019}
\usepackage{url} 
\usepackage{lineno}
\usepackage[inline]{trackchanges} 
\usepackage{soul}
\usepackage{amsmath}

\draftfalse

\journalname{Geochemistry, Geophysics, Geosystems}

\begin{document}

%
%


\title{A method for measuring and analyzing four-dimensional flow fields in viscous Rayleigh-B\'enard convection}

%
%




\authors{Xiyuan Bao\affil{1}\thanks{Now at the Department of Earth and Planetary Sciences, Harvard University, Cambridge, MA, USA}}

\authors{Carolina Lithgow-Bertelloni\affil{1}}

\affiliation{1}{Department of Earth, Planetary, and Space Sciences, University of California, Los Angeles, Los Angeles, CA, USA}





\correspondingauthor{Xiyuan Bao}{xiyuanbao@g.ucla.edu}



\begin{keypoints}
\item A complete experimental and computational workflow for measuring and analyzing 4D flow in high-Prandtl-number viscous convection.

\item  Combines Scanning Stereoscopic PIV with Lagrangian Coherent Structure and clustering analyses to identify and track plumes objectively.

\item Enables quantitative measurement of plume position, spacing, and morphology, with added constraints from optical distortion.
\end{keypoints}

%
%

%
%


\begin{abstract}
Laboratory investigation of convective flow in viscous fluids is crucial for various engineering and geophysical applications, including vigorous convection in planetary mantles, which involves multiple rising plumes. Many experimental techniques have been developed to visualize the flow and plumes, but most previous measurements remain largely qualitative. Here we present a method to measure and analyze four-dimensional (4D) plume-bearing viscous fluid flow, extending volumetric velocimetry approaches to the high-Prandtl-number laminar regime relevant to mantle convection. First, a customized three dimensional (3D) Scanning Stereoscopic Particle Image Velocimetry (SSPIV) system is applied to a Rayleigh-B\'enard experiment suitable to study the dynamics of Earth's interior. We report the first quantitative 4D velocity measurements of multiple interacting laminar plumes at high Prandtl numbers. The raw velocity data are postprocessed with a Lagrangian Coherent Structure (LCS) and cluster analysis pipeline, tailored for clusters of viscous plumes, which allows quantitative, transport-based material boundary tracking of the plumes in space and time. We show example results to demonstrate the power of our method while fully exploring the dynamics in a companion paper \citeA{bao2025evolution}. Our analysis quantifies the rich dynamical behavior of multiple interacting plumes. We suggest our method of measurement and analysis can help better quantify the morphology, interaction, and evolutionary pathways of plumes feeding Earth's volcanic hotspots and other planetary interiors.
\end{abstract}

\section*{Plain Language Summary}
Convection in viscous fluids, where hotter material rises and cooler material sinks, plays a central role in shaping planetary interiors, including Earth's mantle. Hot, buoyant structures called plumes transport heat and material from deep within Earth to the surface, but details of their shape, motion, and interaction remain difficult to measure. This study presents a new method to directly observe and analyze plume-bearing convection in the laboratory, capturing full 3D flow over time (a 4D velocity field) in a high-viscosity fluid relevant to mantle conditions. Using a specialized particle image velocimetry system and advanced analysis techniques, we measure how plumes emerge, evolve, and interact in space and time. We also analyze the plumes in a reference frame moving with the fluid to identify boundaries and behavior of individual plumes objectively. The distortion of light rays due to plumes also helps constrain plume head and stem size. Our results reveal that plumes are more dynamic than previously thought, often merging or changing shape as they rise. This new method provides a powerful tool for understanding how plumes operate deep inside planets, with implications for interpreting volcanic hotspots and the internal evolution of Earth and other rocky bodies.

\section{Introduction}\label{sec:intro}

A quantitative description of the morphology and dynamics of convective features in viscous fluids is important due to its diverse applications in engineering and geoscience as explored across experimental fluid mechanics and geodynamics \cite{ciofalo2003tomographic, limare2013microwave, walbecq2024fully}. It can be used to design glass furnaces \cite{chiu2008very}, engine lubricants \cite{renon2021experimental}, and the cooling core of nuclear reactors \cite{holzbecher2021parameter}. Viscous convection is also relevant to the dynamics of salt domes \cite{hudec2007terra}, planetary ices \cite{greve2006fluid} and planetary interiors \cite{chandrasekhar1953xxv}. Plumes, localized regions of fast releasing buoyant fluid, are common in vigorous viscous convection. Multiple laminar plumes are expected to present in the convection of Earth's and other planetary and exoplanetary mantles, basally heated by the cooling of their liquid iron cores, and cooled from above by space, or atmospheres and oceans. Plumes are the source of many volcanic hotspots \cite{Morgan1971}, providing important geochemical tracers of deep mantle evolution \cite{zindler1986chemical}. 

Plumes in the mantle may not resemble the idealized isolated plumes originally envisioned or imagined. Plume spacing may vary widely raising the possibility of plume-plume interactions \cite{pera1975laminar,zhang2023emerging}. Plumes may also not survive all the way from the Core-Mantle Boundary (CMB) to surface intact, but instead may split \cite{Liu2020,zhang2026bilateral} or detach from their base \cite{Davaille2005}. The possibility of such phenomena highlight the need for a rigorous quantitative definition of a plume and its morphology \cite{Davaille2011,Cagney2015}. A better understanding of merging, splitting, and detachment is essential for a more robust interpretation of geochemical tracers.

Our challenge is to measure the full 4D velocity field of a naturally evolving multi-plume system and rigorously identify the plumes and their behavior with plume material boundaries and entrainment pathways. Quantitative plume identification and tracking have a substantial history in numerical studies, using temperature fields \cite{labrosse2002hotspots, Arnould2020}or combined thermal and Lagrangian diagnostics \cite{Farnetani2002,Lin2006a,Samuel2014}. What remains out of reach, however, whether in simulation or experiment, is resolving the material-transport details of multiple coexisting, interacting plumes at the resolution required to track entrainment, because the necessary physical scales are unknown a priori and often prohibitively costly \cite{leng2023progress,mohr2023challenges}.

Laboratory study of convection in viscous fluids with plumes can be traced back to early salt dome experiments \cite{nettleton1934fluid}. Since then, much experimental work has focused on obtaining scalings related to the steady-state structure and the life cycle of an isolated plume, or a cluster of plumes \cite<see>[and references therein for early and more recent contributions]{whitehead1975dynamics,Lister1989,Griffiths1990,Daville2015treatise}. A variety of techniques have been adopted to visualize the development of laminar plumes in viscous fluids. Some are predominantly qualitative morphological visualizations, such as dye \cite{whitehead1975dynamics}, shadowgraphs \cite{shlien1976some}, or differential interferometry \cite{kaminski2003laminar}, which capture plume shape and evolution but do not directly yield calibrated velocity or temperature fields. Others are quantitative but spatially restricted, including single-plane thermochromic liquid crystal thermometry \cite{Lithgow-Bertelloni2001}, planar laser-induced fluorescence \cite{kumagai2007fate}, and planar Particle Image Velocimetry \cite{Davaille2005}, which deliver calibrated fields within a selected cross-section. For a single axisymmetric plume from a point heat source, a 2D plane through the plume axis can in principle recover its 3D structure by symmetry. Resolving the time-dependent holistic view of multiple interacting plumes and their surroundings in a system without such symmetry, however, requires volumetric measurements. There are some earlier volumetric PIV efforts such as scanning PIV \cite{dong1990piv}, scanning stereoscopic PIV \cite{brucker1996}, tomographic PIV \cite{elsinga2006tomographic}, however they tend to address lower-viscosity or turbulent flows. \citeA{ciofalo2003tomographic} introduced scanning PIV in viscous convection experiments with a single camera, but no out-of-plane velocity could be obtained, resulting in a 2.5D velocity in space. A similar setup was also used in \citeA{Androvandi2011} and \citeA{limare2013microwave}. Meanwhile, \citeA{Newsome2011} and \citeA{Cagney2015} constructed a 3D Scanning Stereoscopic PIV (SSPIV) system with two cameras, and for the first time, measured the 4D velocity (i.e. 3D velocity in space, and 1D in time) for an axisymmetric single thermal plume from spot heating in a viscous fluid. 
\citeA{walbecq2024fully} recently performed 3D Particle Tracking Velocimetry for Rayleigh-B\'enard systems, but focused on lower viscosity fluids (Prandtl number $Pr<100$) and did not analyze plumes. The inertia dominates the flow with $Pr<1$, and lagging of flow development is obvious with intermediate $Pr$ (1~100), while the behavior is very similar to no inertia regime with high $Pr$ of 1000 or more \cite{whitehead2013numerical}. To the best of our knowledge, no 4D velocity has ever been obtained in the laboratory for multiple interacting plumes in effectively inertia free ($Pr\ge1000$)  viscous convection. The lack of such quantitative measurements limits the analysis of  plume dynamics and material transport in high $Pr$ convection relevant to planetary mantles. A comparison with previous work can be found in Table~\ref{tab:previous_work}.

\begin{table}
\label{tab:previous_work}
\small
\setlength{\tabcolsep}{2.5pt}
\renewcommand{\arraystretch}{1.2}
\centering
\begin{tabular}{p{2 cm} p{2cm} p{2cm} p{1.4cm} p{3cm} p{2.5cm}}
\hline\hline
\textbf{Study} & \textbf{Method} & \textbf{Regime} & \textbf{Dim.} & \textbf{Key Contribution} & \textbf{Limitation vs. Present Methodology} \\
\hline
\citeA{dong1990piv} & Scanning PIV & Turbulent flow & 2.5D +time & First volumetric PIV attempt & Not viscous convection; no out of plane velocity \\
\citeA{brucker1996} & Scanning Stereoscopic PIV & Turbulent flow & 4D & First 4D velocity with PIV &  Not viscous convection\\
\citeA{ciofalo2003tomographic} & Scanning PIV & Viscous Rayleigh-B\'enard convection ($Pr\sim10^4$) & 2.5D +time & First scanning volumetric PIV in viscous convection & No out of plane velocity; low convective vigor ($Ra\sim10^4$) \\
\citeA{elsinga2006tomographic} & Tomographic PIV & Turbulent flow & 4D & Simultaneous 4D velocity measurement, high seeding density  & Not viscous convection \\
\citeA{Androvandi2011} &  Scanning PIV & Viscous convection ($Pr>>1$) & 2.5D +time & Wide range of viscosity contrast explored & No out of plane velocity; analysis focused on selected planes \\
\citeA{limare2013microwave} &  Scanning PIV & Viscous convection ($Pr>10^3$) & 2.5D +time & First scanning volumetric PIV in viscous convection with internal heating & No out of plane velocity; only 1/3 volume measured \\
\citeA{Newsome2011,Cagney2015} & SSPIV for single plume & Viscous flow, single-spot heating ($Pr>10^3$) & 4D & First 4D viscous plume velocity field & Axisymmetric single plume; small domain \\
\citeA{walbecq2024fully} & Tomographic PTV & Viscous convection ($Pr<100$)& 4D & Special treatment of the near-wall flow & moderate $Pr$ (8 to 85); lower seeding density \\
\textbf{This study} & \textbf{SSPIV} & \textbf{Viscous convection ($Pr>10^3$)} & \textbf{4D} & \textbf{First end-to-end 4D methodology for multiple viscous plumes} & --- \\
\hline\hline

\end{tabular}
\caption{Summary of previous and present studies on volumetric velocimetry in different flow regimes, applicable to viscous plumes.}
\end{table}

Different visualization methods can lead to different definitions for the boundary of plumes \cite{Davaille2011,Newsome2011,Cagney2015}, because they target different physical aspects of a plume, its thermal anomaly, its dynamic signature, or the material it transports, and the choice of definition therefore depends on the physical question being asked. Thermal boundaries can be drawn from isotherms or local temperature gradients \cite{labrosse2002hotspots,zhong2005dynamics,Leng2012}, which are appropriate for comparison with seismic tomography. Dynamic boundaries can be drawn from iso-velocity contours, local velocity gradients, radial-velocity or heat-advection gradients, stagnation points, or streamlines \cite{Davaille2011,Hassan2015,Arnould2020,shevkar2022separating}, which tracks the flow from a snapshot. Different criteria are complementary and in general do not identify identical boundaries \cite{Davaille2011,Cagney2015}; the ambiguity in plume boundary definition may result in differing interpretations of the detection, size, morphology, composition, and strength of the plume \cite{labrosse2002hotspots,zhong2005dynamics,kumagai2008mantle}, changing our understanding of the energy budget and the physical and chemical state of Earth's and other planetary mantles \cite{farnetani2009dynamics,hoggard2020hotspots}. Lagrangian Coherent Structure (LCS) theory provides a complementary, transport-based definition: a mathematically well-defined material surface that cannot be crossed by the flow and whose identification is based on accumulated flow history instead of a specific temperature or velocity threshold \cite{haller2000finding,shadden2005definition,Shadden2011,Haller2015,hadjighasem2017critical}. Lagrangian analysis uses metrics in a frame moving with the fluid and can extract the material boundary of plume heads and stems, advancing our understanding of material entrainment in mantle convection and planetary interior dynamics \cite<e.g.,>{ferrachat1998regular,Farnetani2002,Lin2006b,Samuel2012,Cagney2015,thomas2024mixing}. As we will show, the resulting material boundary appears as a sharp ridge in the corresponding scalar field, considerably sharper than smooth isocontours like for velocity, so the identified boundary is less sensitive to the specific threshold value. Lagrangian analysis is therefore the natural diagnostic to characterize plume material transport, complementary to thermal and dynamic criteria used for other questions. Applying such analysis to viscous convection requires time-resolved 4D velocity data, precisely the data needed to establish a physical basis for surface-to-source mapping of chemical anomalies to a geographical location in the mantle.

Here we describe our methodology for meeting this challenge, namely measuring and analyzing the full 4D velocity field of multiple simultaneously evolving thermal plumes in a high-Prandtl-number fluid, which has not previously been achieved. Our approach combines SSPIV mapping of the 4D velocity field with Lagrangian analysis to identify plumes, in a basal heating experiment that produces multiple plumes, rather than the spot heating \cite{Cagney2015} that was used to produce a single isolated plume. This paper builds on previous work in which we focused on SSPIV \cite{Newsome2011,Cagney2015,Cagney2016a,Cagney2016b}, and solving one particular problem in volumetric PIV in viscous fluids: that of optical distortion \cite{bao2024self}. Further geophysical implications of our results are presented in a companion paper \cite{bao2025evolution}. In this paper, we briefly describe the experimental setup (Sec. \ref{sec:exp}) and highlight the analysis methodology necessary to obtain the fluid flow, the distribution of plumes (Sec. \ref{sec:analysis}) and their evolution. Examples of the results are shown in Sec. \ref{sec:rslt1}. Finally, in Sec. \ref{sec:discussion}, we briefly discuss our approach in comparison to previous studies and how it helps us reveal the rich dynamics of plumes. Detailed analysis of the plume flow, including the network of possible evolutionary pathways, scaling and geophysical interpretation are presented in a companion paper \cite{bao2025evolution}.

The end-to-end methodology from experimental measurements to readily available plume statistics and evolution requires integrating multiple components that need elaboration and validation. Hence, we have chosen to present it in this standalone paper rather than as an appendix or supplementary material. In addition, in the planetary and solid Earth fluid dynamics communities, experiments are rare compared with numerical simulations, and often undervalued or dismissed. By presenting an end-to-end methodology also applicable to numerical simulations, along with new measurements of plume dynamics, we hope to show the power of combining experiments and analysis to further quantitative analysis of plume flow. In the companion study \cite{bao2025evolution}, we apply this methodology to quantify the collective evolution of plumes, demonstrating its broader geodynamic relevance.

Compared with prior volumetric and scanning velocimetry, the present system uniquely resolves 4D velocity fields in a high-Prandtl-number regime, bridging laminar laboratory convection and mantle-scale numerical models (Table~\ref{tab:previous_work}). By revealing the planform of convection, locating the distribution of plumes and evolutions at different scales, our detailed measurements and tailored methodology can help unlock the full potential of laboratory experiments, which operate on real-world physics with infinite resolution.

\section{Laboratory experiment}
\label{sec:exp}
The general experimental setup was described in \citeA{bao2024self}. Here, we only summarize the key SSPIV components, and details not covered previously, including image postprocessing and data analysis. For completeness, we also include more information about the experiment in Appendix \ref{app:details}.

\subsection{Test section}
\label{sec:tank}
\begin{figure*}
  \centering
  \includegraphics[width=\linewidth]{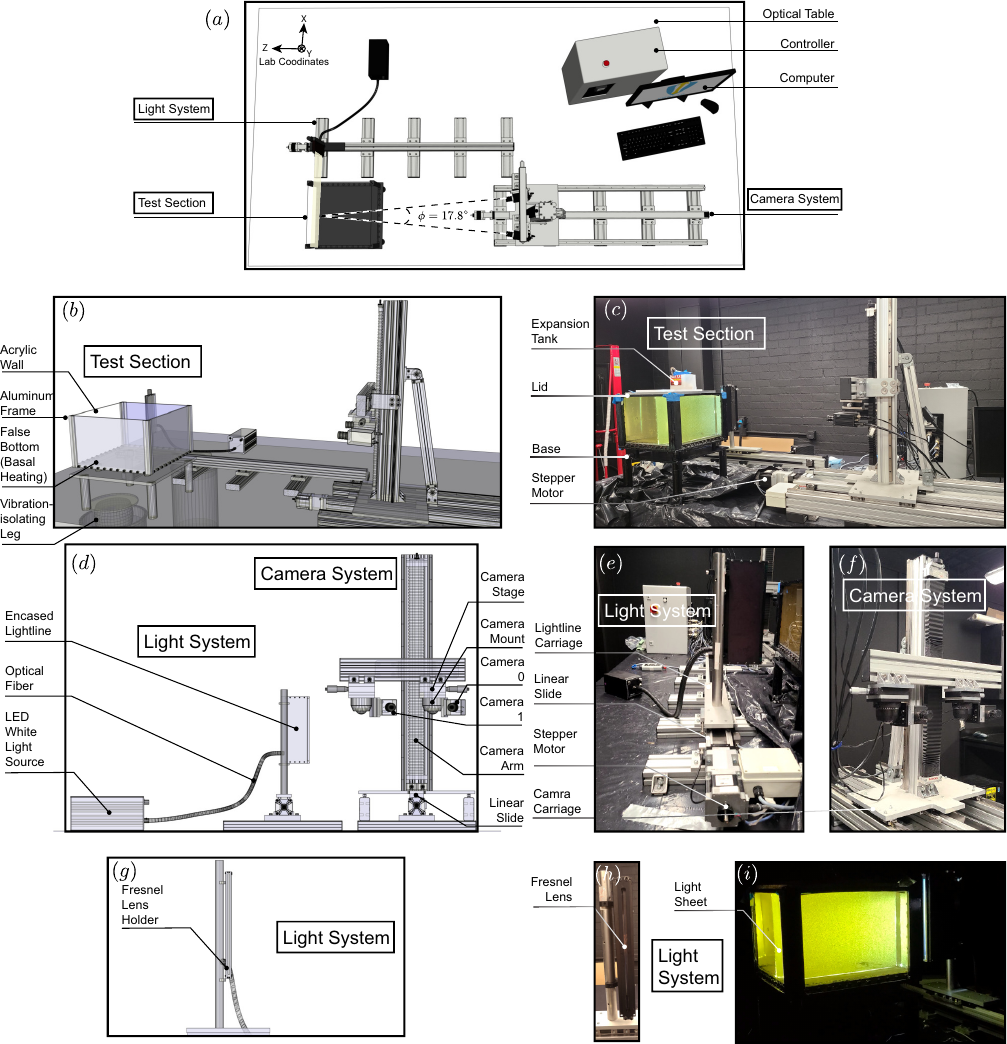}
  \caption{Experimental setup for the 3D Scanning, Stereoscopic, Particle Image Velocimetry (SSPIV) system. (a) Top view. Monochrome renderings of the model are shown in (b), (d) and (g) from front, left, and zoomed in to the front of the lightline. Lab photos of the system correspond to the sketches (c for b, e and f  for d, and h for g). (i) A photo of the tank during the experiment without environmental light.
  }
\label{fig:lab}
\end{figure*}

The experiment was carried out in a plexiglass tank, with internal dimensions 405 mm $\times$ 275 mm $\times$ 405 mm along $x,y,z$ and 10 mm thick wall (Fig. \ref{fig:lab}a, b, c, Appendix \ref{app:wall}). The tank lid was maintained at room temperature (25 $^\circ$C), while the bottom was heated using a silicone heating mat regulated by a PID controller. At $t=0$ s, the heater was turned on, and in 250 s, the bottom temperature increased almost linearly from 25 $^\circ$C to 80 $^\circ$C, which is much shorter than the relatively transient stage ($>3500$ s, \cite{bao2025evolution}). Then the bottom temperature was kept constant until the end of the experiment ($\sim$ 8000 s).

\subsection{Working fluid}
\label{sec:working_fluid}
The strength of thermal convection in viscous fluids can be described by the global Rayleigh number $Ra$, defined as:
\begin{eqnarray}
  Ra = \frac{\tau_\mathrm{diff}}{\tau_\mathrm{conv}} = \frac{H^2/\kappa}{\frac{H}{\rho \alpha \Delta T g H^2/\eta}} = \frac{\rho g \alpha \Delta T H^3}{\eta \kappa}
  \label{eqn:Ra}
\end{eqnarray}
where $\rho$ is the reference density, $g$ is gravity, $\alpha$ thermal expansivity, $\Delta T$, the temperature difference driving the flow, $H$ is the thickness of the convective region, $\eta$, the dynamic viscosity and $\kappa$ , the thermal diffusivity. $Ra$ describes the ratio of the diffusive over convective timescales for heat transport.

The effect of inertia can be represented by the Prandtl number $Pr$, defined as:
\begin{eqnarray}
  Pr = \frac{\eta/\rho}{\kappa} = \frac{\nu}{\kappa}
  \label{eqn:Pr}
\end{eqnarray}
where $\nu$ is the kinematic viscosity. The effect of inertia is essentially negligible for very viscous fluids ($Pr>$1000),  \cite{whitehead2013numerical}.

In  our experiment, a very viscous fluid ($Pr>3000$ at the maximum temperature in the experiment, essentially inertia-free) was heated from below, undergoes vigorous convection  at $Ra=1.9 \times 10^6$. This is a relatively high Rayleigh number similar to Earth's mantle and $>>$ than  Rayleigh critical, $Ra_\mathrm{c}$,  (1708 \cite{Jeffreys1928}). The fluid flow is dominated by thermal plumes. The working fluid was Gateway Du-Crose 3 63/43 corn syrup. It is clear and chemically stable. Its viscosity is high and strongly temperature-dependent (super-exponential), measured as:
\begin{equation}
  \eta = \mathrm{exp}(4.642 \times 10^{-4} T^2 - 1.246 \times 10^{-1} T + 6.325)
  \label{eqn:visc}
\end{equation}

Here temperature $T$ is in $^\circ$C, viscosity $\eta$ is in Pa$\cdot$s. With temperatures between 25 to 80 $^\circ$C the maximum viscosity contrast ($\gamma=\eta_\mathrm{max}/\eta_\mathrm{min}$) was 65. The temperature-dependency of density $\rho$ and refractive index $n$ was also measured and compared with manufacturer references (cf. Appendix \ref{app:fluid}). The thickness of the bottom thermal boundary layer is approximately 7.5 mm, and the characteristic plume rising speed is 0.5 to 1 mm/s (cf. \cite{bao2025evolution}). Table \ref{tab:syrup} summarizes all the properties of the fluid measured in the lab or obtained from the manufacturer at room temperature.

\subsection{Velocity measurements}

\subsubsection{SSPIV system}
The SSPIV system used in our experiment was based on the initial design by \citeA{Newsome2011,Cagney2015}, and improved to scan a larger test section, improve positioning precision, camera resolution, and operating software. It was built by \citeA{Cagney2016b} at University College London, then reassembled and further updated at UCLA (Fig. \ref{fig:lab}). The system sits on a new optical table (Newport RPR-510-12/SL-600-423.5) for vibration isolation. Two new higher resolution cameras are mounted on a movable arm, at the height of the center of the test section, in a stereoscopic setting with a separation angle of 17.8$^\circ$ (JAI AT-200-CL CCD with Fujinon TF15DA-8 15 mm lenses, Fig. \ref{fig:lab}d, f). A new white LED light source (Hecho S5000) is connected to a Volpi fiber optic lightline with a cylindrical lens (Fig. \ref{fig:lab}d, e). A Fresnel lens is mounted in front of the cylindrical lens to produce a non-divergent, thin (5mm) light sheet illuminating a plane section of the fluid (Fig. \ref{fig:lab}g, h, i). The stereoscopic arrangement and the finite thickness of the light sheet allow the out-of-plane velocity to be extracted. The system can be extended to include thermochromic liquid crystals to measure temperature \cite{Newsome2011,Cagney2015}. This is possible because we use a white light source rather than a laser, and 3-chip CCD cameras that isolates RGB components individually, which can then be used to calibrate from hue to temperature \cite{dabiri1991digital}, letting us access a range of temperatures, usually a few degrees \cite{Newsome2011,Cagney2015}, which can be extended to 30 degrees with a spectroscope, \cite{toriyama2016new},  not only individual isotherms. The room is completely dark during the experiment to avoid stray reflections (Fig. \ref{fig:lab}i).

The cameras and lightline are mounted on motion-controlled linear slides with pitch screws, reaching repeatable precision positioning of $\sim5$ microns \cite{Bao2024PhD}. (Fig. \ref{fig:lab}a, d, e, f). The data acquisition process involves synchronously driving the cameras and the lightline using separate stepper motors with drives from Applied Motion Products. Synchronicity is achieved via the PC clock, which has ms level precision. A computer with an in-house code SPIVET-Control \cite{Newsome2011} generates the motion signal and sends it to the controller through the drives to the stepper motors. We use the National Instruments PCIe-1429 frame grabber for image data acquisition from the cameras. SPIVET-CONTROL uses APIs from National Instruments to control camera exposure and saving the images to the disk.

We use 30 ppm silver-coated polymer spheres as tracer particles (Potters Conduct-O-Fil SP30S20, diameter distribution 30.5$\pm3$ $\mathrm{\mu m}$). This tracer is highly reflective and neutrally buoyant (density 1487 kg/m$^3$, versus 1427 kg/m$^3$ of the fluid). This concentration allows us to obtain effectively 7~8 tracers per 16 pixel x16 pixel window.

\subsubsection{Image data}
\label{sec:image_data}
\begin{figure*}
  \centerline{\includegraphics[width=\linewidth]{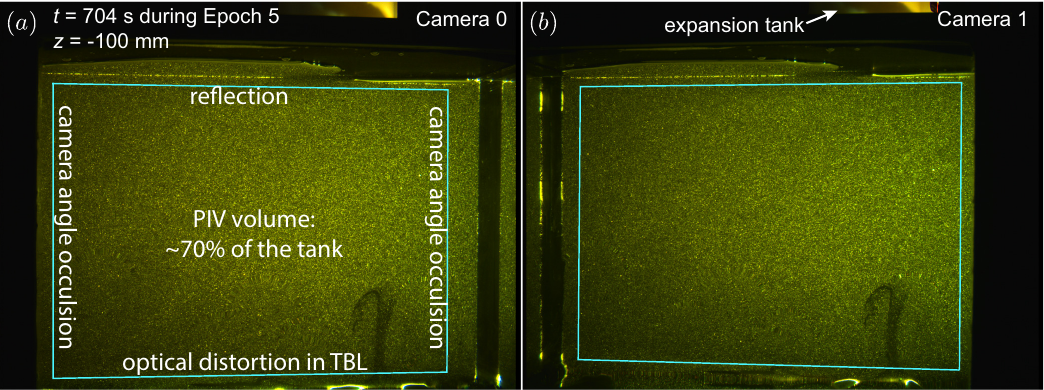}}
  \caption{Particle images at 704 s during Epoch 5 at $z=-100$ mm from Camera 0 (\textit{a}) and Camera 1 (\textit{b}). The cyan box outlines the SSPIV region in the in-plane space. Reasons for excluding regions shown along the box sides for (\textit{a}).}
\label{fig:stereo}
\end{figure*}

We scanned 69 tank planes at 5 mm intervals along the $z$ axis (the directions of the lab coordinate system are shown in Fig. \ref{fig:lab}a) and acquired 60 epochs (Epoch 0 to 59) of data, with no pause between consecutive epochs. Each epoch had a duration of $\sim$130 s (cf. \ref{app:FTLE_width} for discussion on the limitations our temporal resolution imposes), so the time between the start of two consecutive epochs is also $\sim$130 s. Two frames are acquired for each plane in an epoch, for a total of 16,560 images for all epochs. The tank was not scanned in its entirety in each epoch, rather 2 planes were acquired before that section of the tank was rescanned, so the interframe interval (i.e., the time between the two frames acquired at the same plane position) was $<1.5$ s, adequately short relative to the flow timescales in the experiment. The time between two consecutive plane positions is variable due to the two-plane-in-a-group scanning, and it is either near 0.5 s or 3 s. An example of the raw image data from both cameras is shown in Fig. \ref{fig:stereo}. The images (1648$\times$1226 pixels) are then corrected with photogrammetric calibration (processed and cropped as 1837$\times$1247 pixels), which maps the image coordinates to lab coordinates on a calibration target considering a pinhole camera model \cite{Newsome2011}. The corrected image of Fig. \ref{fig:stereo}a is shown in Fig. \ref{fig:raw}a.

\subsubsection{SSPIV data processing}
\label{sec:SSPIV_process}
The velocity field was obtained with another bespoke internal code, SPIVET-UCLA, an upgraded version of SPIVET \cite{Newsome2011}. SPIVET-UCLA is designed to work with SSPIV image data and its newly added hybrid CPU/GPU parallel processing which compresses post-processing time from weeks to hours. Cross-correlation and WiDIM \cite{scarano2000advances,Newsome2011} algorithms were used, followed by optical flow and thin-plate-splines to enhance multi-scale performance\cite{Liu2009,Newsome2011,Bao2024PhD}. The final vector resolution is $3.53 \times 3.53$ mm \cite<extendable to 0.2 mm, >{bao2024self} in-plane and 5 mm inter-plane.

SSPIV uses the common region covered by the two cameras. Due to the occlusion from the edge (frame) of the tank (Fig. \ref{fig:stereo}), the area near the wall along the $x$-axis was not included. The scanning along the $z$ direction was again limited by the occlusion from the frame of the tank. We also removed about 5 mm at the top and bottom of the fluid domain to avoid the possible reflection and systematic optical distortion in the thermal boundary layer (TBL). The final volume covered by SSPIV along $x-y-z$ was [-13.2, 336.0], [2.7,256.7], [-342.5,2.5] mm, or in percentage $>86\%$, $>92\%$ and $>85\%$, respectively ($68\%$ fluid volume). The origin, near the upper left corner of the furthest plane when viewing from the cameras, was determined during photogrammetric calibration \cite{Bao2024PhD}. For reference, the upper left corner far from the cameras of the fluid domain is at (-26.6,-1.9,32.5), while the center of the fluid domain is at (175.9, 135.6, -170.0).

\subsubsection{Velocity post-processing: Removing optical distortions and synchronizing velocities}
\label{sec:velocity_post_process}

\begin{figure*}
  \centerline{\includegraphics[width=\linewidth]{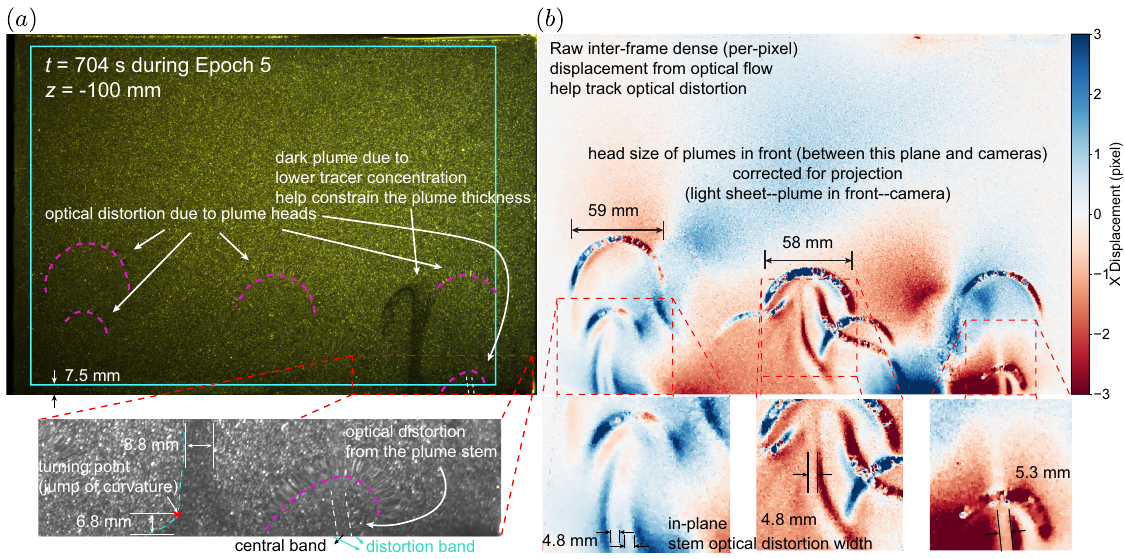}}
  \caption{A particle image (\textit{a}) and corresponding in-plane inter-frame displacement (\textit{b}) at 704 s during Epoch 5 at $z=-100$ mm for our experiment. (\textit{a}) The image is from camera 0, and has been dewarped as if the camera were perpendicular to the center of the tank. The cyan box shows the SSPIV region in in-plane space, excluding the bottom 7.5 mm corresponding to the thermal boundary layer thickness. The visible dark plume on the lower right is due to the entrainment of poorly mixed material with fewer tracers in and above the boundary layer near the bottom. The width of the dark plume stem is 8.8 mm, while the height of the turning point (jump of curvature, with the derivative of curvature changing sign, from the root to the stem) is 6.8 mm, as shown by the red star (plume outlined by the blue dashed line), in the zoomed inset outlined by the red rectangle. In the same inset, the head (pink dashed line) and stem (white dashed line, highlighting ``central band" as in Fig. \ref{fig:ray-tracing}) of a plume between the illuminated plane and cameras (``plume in front", or $p_\mathrm{if}$) next to the dark plume can also be identified from optical effects (Appendix \ref{app:distortion}). (\textit{b}) The raw horizontal displacement affected by the $p_\mathrm{if}$, obtained with optical flow  \cite{Brox2004,Liu2009} for every pixel ($0.22\times0.22$ mm). Each pixel of displacement is equivalent to 0.15 mm/s of velocity. See Appendix \ref{app:distortion} for more analysis. The sizes of the plume head (corrected for projection) and the plume stem optical distortion from plumes in front are also shown. The dark plume does not cause artifacts in the final PIV results. This figure shows how we can quantitatively extract the size of the plume head and stem. See \citeA{bao2024self} figure 10 for shadow-free examples of the raw image, optical flow and PIV results.}
\label{fig:raw}
\end{figure*}

We observed obvious optical distortion from plumes between the illuminated plane and the cameras, which corrupted the extracted velocity of almost all planes (e.g., Fig. \ref{fig:raw}b), and might have prevented previous attempts at accurate 4-D PIV in convection experiments\cite{elsinga2005evaluation}. We have designed a fast and accurate custom filter based on non-local means\cite{buades2011non} and morphological operations\cite{serra1983image} to mitigate this problem (cf. \citeA{bao2024self} for more details). We treat the optical distortion not only as a nuisance, but as a tool to provide direct constraints on the size of the plumes\cite{bao2024self}, such as the width of the plume stem (cf. section \ref{sec:thickness_method}).
 
We synchronized the velocity at each plane in the same epoch by interpolating to the time of the last plane with cubic splines. The 3D velocity at the end of each epoch (from Epoch 0 to 58) was obtained, except for the last epoch (Epoch 59), which required extrapolation.

The dominant source of uncertainty in the synchronized 4D velocity field is the temporal interpolation between scans of neighboring planes, that is, the cubic-spline interpolation over the $\sim$130 s inter-epoch interval (cf. Appendix~\ref{app:FTLE_width}). Other sources are individually small relative to this: stage positioning is repeatable to a few microns (cf. ~\ref{app:pos_calibration}); the photogrammetric calibration introduces a $\sim2$ pixel (a pixel is 0.22 mm) localization offset on average (cf. Appendix~\ref{app:photo_calibration}); and the per-plane PIV cross-correlation with the optical-distortion filter achieves sub-pixel displacement accuracy (cf.\citeA{bao2024self}). The out-of-the-plane velocity reconstructed from the SSPIV setup has an error on the same order of magnitude of that from the in-plane velocity (and could be as much as four times larger). The temporal interpolation, by contrast, is unavoidable in any scanning velocimetry of a time-evolving multi-plume flow, and it is the primary source of uncertainty that propagates into Lagrangian analysis of the velocity field (cf. Appendix~\ref{app:FTLE_width}). The uncertainty quantification of the synchronized 4D velocity field requires a digital twin of the experiment and we will explore this in the future.

\section{Analysis methods}
\label{sec:analysis}
\begin{figure*}
  \centerline{\includegraphics[width=\linewidth]{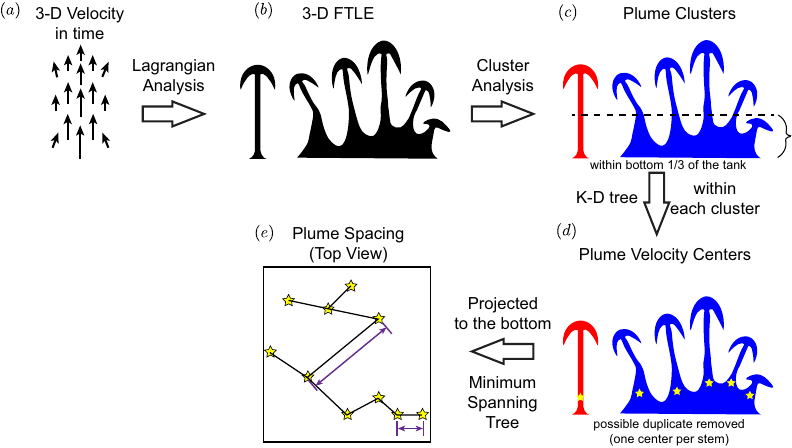}}
  \caption{Schematic of the analysis methodology based on SSPIV observations. The 4D, or 3D velocity field in time indicated by the vectors in (a) can be used to generate the Finite-Time Lyapunov Exponent (FTLE) field in (b) using Lagrangian analysis. With cluster analysis, the ridges (local maxima) in the FTLE that correspond to plume material can be separated into individual clusters in (c) as point clouds. Within each cluster (point cloud), the plume center(s), or upwelling velocity maxima, can be located using the K-D tree, shown as stars in (d). Plume centers are then projected onto the tank bottom. The minimum to maximum (purple arrows) plume spacing can be found using the minimum spanning tree in (e) as black lines connecting the stars.}
\label{fig:analysis_schematic}
\end{figure*}

The analysis methodology based on the observed velocity is summarized in Fig. \ref{fig:analysis_schematic}. We characterize all plume material in the flow to isolate each plume in space. We also use the particle images themselves to extract direct information about the size of the plumes.


\subsection{Lagrangian analysis}
\label{sec:lagrangian_method}
While our PIV velocity data is in a fixed spatial (Eulerian) frame, Lagrangian analysis allows us to extract information about fluid packages interacting, stretching, and stirring in a moving Lagrangian frame. For example, we can find Lagrangian Coherent Structures (LCSs), or 3D material surfaces that cannot be crossed by the flow, and separate relatively distant regions \cite{Shadden2011,Haller2015}. Lagrangian analysis has been applied to mantle convection but has been limited to 2D numerical simulations \cite{ferrachat1998regular,Farnetani2002,Farnetani2003,Lin2006b,Obermaier2011,Samuel2012,Samuel2014} or experiments focused on a single thermal plume \cite{Newsome2011,Cagney2015,Cagney2016a,Cagney2016b}.

To perform Lagrangian analysis, passive tracers are typically advected numerically through the measured velocity field over a prescribed interval of time. There are repelling LCSs, where local fluid parcels tend to advect away from them, and attracting LCSs, where local fluid particles tend to advect towards them. The LCSs can be approximated by ridges (local maxima) in the Finite-Time Lyapunov Exponent (FTLE) field $\sigma_f$:
\begin{equation}
  \sigma_f = \frac{1}{2|(t-t_0)|}\mathrm{log} \left( \mathrm{max} \left( \frac{||\delta \boldsymbol{x}(t)||}{||\delta \boldsymbol{x}(t_0)||} \right) \right)
  \label{eqn:FTLE}
\end{equation}
where $t_0$ and $t$ are the beginning and end of tracer advection, respectively; $\delta \boldsymbol{x}$ is the distance of neighboring particles originally located at $\boldsymbol{x}$.  The FTLE field is therefore a measure of maximum exponential separation (or stretching) between neighboring particles from $t_0$ to $t$. The stretching term $\mathrm{max} \left( \frac{||\delta \boldsymbol{x}(t)||}{||\delta \boldsymbol{x}(t_0)||} \right)$ can be computed from the largest eigenvalue of the right Cauchy-Green tensor $\boldsymbol{M}$:
\begin{equation}
  \boldsymbol{M} = (\nabla\boldsymbol{F})^T\nabla\boldsymbol{F}
  \label{eqn:CG}
\end{equation}
where $\boldsymbol{F}$ is the flow map that maps tracer locations from the beginning (e.g.,$\boldsymbol{x}_0,t_0$) to the end of its trajectory (e.g.,$\boldsymbol{x}_1,t_1$), i.e., 
\begin{equation}
  \boldsymbol{F}(\boldsymbol{x}_0,t_0) = \boldsymbol{x}_0; \boldsymbol{F}(\boldsymbol{x}_0,t_1) = \boldsymbol{x}_1
  \label{eqn:flowmap}
\end{equation}

The FTLE field can be computed either forward in time, in which the ridges correspond to repelling LCSs; and backward in time, by reversing the sign of velocity and time. We focus on the backward-time FTLE in this paper because the FTLE ridges (high $\sigma_f$) are attracting LCSs and, in our experiment, correspond to the plume boundaries, including head and stem \cite[Fig. \ref{fig:analysis_schematic}a, b]{Cagney2015}. For simplicity, we use $\sigma_f$ to represent the backward-time FTLE field for the rest of the paper.

To compute the backward-time FTLE field, we initialized tracers from each of the later epochs (Epoch 3 to 58) and advect them back to Epoch 0 with the corresponding time-span ranges from 3 to 58 epochs. Following the standard Lagrangian-based FTLE calculation procedure \cite{shadden2005definition}, tracers are released on a uniform grid at the final time $t_1$ and integrated backward to $t_0$, so that the resulting flow map and FTLE field inherit the uniform spatial resolution of the release grid rather than the evolving local density of the advected tracers. This allows us to obtain a series of $\sigma_f$ fields from the first epoch that includes plume initiation (Epoch 3) until the end of the experiment. The amplitude is comparable as $\sigma_f$ has been normalized by the time-span. Tracers ($>$17 million) were uniformly distributed at initiation in each plane ($4\times4$ tracers along $x$-$y$ for each PIV vector, and 2 tracers along $z$, so 32 tracers in total around each PIV vector), leading to a resolution of $0.88 \times 0.88 \times 2.5$ mm for the FTLE field. We used a fourth-order Runge-Kutta \cite{Kutta1901} scheme with cubic interpolation in space and time to advect the tracers. Some tracers were advected out of the computation (SSPIV) domain because of the incomplete coverage near the boundaries (cf. section \ref{sec:SSPIV_process}), and we used their location when they left the domain to compute $\sigma_f$.

Integrating from each later epoch back to $t=0$, rather than over a fixed sliding window, preserves the complete material-transport history of every plume regardless of when it initiated, which is appropriate for the asynchronously initiated multi-plume flow studied here, where a sliding window would inevitably truncate the early history of any plume that began before the window opened. By construction, $\sigma_f$ is normalized by the integration time, so cumulative stretching from older, no-longer-active events is automatically faded as the integration window grows, while the newest active plume continues to add stretching and remains sharp in $\sigma_f$. The relatively limited temporal resolution of our velocity acquisition further weakens any residual background stretching, since fine-scale stirring is not fully resolved in the underlying velocity field and therefore does not concentrate into sharp ridges that could be confused with active plume signatures. A fixed-window approach can be advantageous in settings where focusing on only the freshest plumes is the center of analysis.

The FTLE ridges, or high $\sigma_f$ highlight the structure of upwelling plumes. We can then objectively draw the material boundaries of each plume with a corresponding $\sigma_f$ threshold. We chose a single conservative threshold ($\sigma_f\ge$0.005 s$^{-1}$) for the entire experiment rather than a series of thresholds at different times or for individual plumes (we note that $\sigma_f$ would change under different experimental conditions). The threshold value is derived directly from the $\sigma_f$ percentile distribution across the experiment, as the minimum of the 95th percentile across post-onset epochs; it sits in a stable gap between the background floor (80th percentile) and the plume ridges (95th-99th percentiles, cf. \ref{app:FTLE_width}). This small constant threshold captures as many plumes as possible.

Because of the limitations on temporal resolution of the velocity measurements mentioned above, the FTLE ridges are wider than desired (cf. section \ref{app:FTLE_width}) which hinders even more detailed plume width characterization. In \citeA{Newsome2011,Cagney2015}, the plume rising speed is up to one order slower than ours, but the scanning speed is the same (i.e., 130 s per epoch scaled to 69 planes), with a  smaller test section (30 vs 40 cm) and fewer planes (38 vs 69 here) to scan. Hence the same plume can be sampled across many more inter-epoch intervals before traversing the test section. The temporal resolution of the FTLE field is therefore set primarily by the faster multi-plume, larger-volume configuration of the present experiment rather than by an intrinsic limitation of the methodology. We emphasize that our methodology would capture the plume width precisely with a better temporal resolution in the FTLE field, as in \citeA{Newsome2011,Cagney2015}. 

\subsection{Cluster analysis}
\label{sec:cluster_method}
The FTLE ridges do not directly provide the numbers of isolated or interconnected plumes unless we separate them by connectivity, in other words unless interconnected plumes are grouped into the same cluster. Note that the very bottom layer of the tank ($<$ 7.5 mm, comparable to the TBL thickness) was not included in the SSPIV domain to avoid optical distortion, due to the strong change of refractive index in the TBL (see section \ref{sec:SSPIV_process}). Therefore the hot ridge-like structures in the TBL which might connect all plumes \cite<e.g.,>{Richter1975} were largely eliminated, and we do not need to further remove the bottom part of the velocity field or FTLE ridges to perform cluster analysis.

We performed cluster analysis via Hierarchical Density-Based Spatial Clustering of Applications with Noise \cite<HDBSCAN,>{Campello2013,McInnes2017}. HDBSCAN can find clusters with high point density in n-dimensional space and isolate them from noise. It builds a hierarchy of clustering with different point densities and is robust with respect to different input parameters (e.g., minimum points in a cluster). We fed the point cloud with $\sigma_f\ge$0.005 s$^{-1}$ in the upwelling region ($U_y<0$, $y$ pointing down) into HDBSCAN. With HDBSCAN, we found a series of clusters, each with one or more plumes, per epoch (Fig. \ref{fig:analysis_schematic}b, c), which we confirmed by visual inspection. Unless otherwise noted, the ``cluster number" associated with each cluster here is not a unique id, and it is not consistent across epochs, but only serves as count of the number of clusters in a particular epoch.

Note that although other cluster analysis methods have been used to analyze Lagrangian coherent structures in fluid flow \cite<e.g.,>{husic2019simultaneous}, their computation is usually intensive for realistic problems. To the best of our knowledge, we are the first to apply a straightforward and scalable cluster analysis method (HDBSCAN) to millions of points and use it to characterize plume statistics and evolutionary pathways in Rayleigh-B\'enard convection.

\subsection{Plume counting}
\label{sec:count_plume}
As each cluster we obtained is usually an aggregation of multiple plumes, we need to further identify each single plume to have a more complete picture of the total number and spatial distribution of plumes. We use the local maximum upwelling velocity ($|U_y|$, or most negative $U_y$) as the center of each plume. We focused on such plume center near the bottom of the tank. For each cluster, we only selected the lower part of the cluster within the bottom 1/3 of the tank (i.e., 91.7 mm), or the entire cluster if its total height is less than this threshold (Fig. \ref{fig:analysis_schematic}c). We searched for the maximum $|U_y|$ within the neighborhood of each point in the selected part of the cluster. In the end, only one velocity center is kept per plume stem (Fig. \ref{fig:analysis_schematic}d).


To implement an efficient local maxima search for the large point clouds (up to $10^6$ points each), we constructed a K-D Tree \cite{Bentley1975} for each cluster (Fig. \ref{fig:analysis_schematic}c, d). This data structure partitions the neighboring points in different levels of subspace and provides fast searching performance.

To analyze the spacing of the plume centers when projected to the bottom, we propose to build a \emph{minimum-spanning tree} \cite{Boruvka1926,Jarnik1930}. This tree is a path that connects all the plume centers using nearest neighbors, and the total length of the path is the shortest, so the minimum spacing among plumes can be acquired. We emphasize that with this method, the average plume spacing is no longer biased by subjective selection, or distant plumes (Fig. \ref{fig:analysis_schematic}d, e).


\subsection{Plume morphology from raw image data}
\label{sec:thickness_method}
As the FTLE ridges are wider than ideal due to limited temporal resolution (cf. Appendix \ref{app:FTLE_width}), we estimate the plume stem width $\delta_\mathrm{m}$, defined by material boundary, directly from the particle images, using two complementary bounds.

\subsubsection{From tracers}
In our experiment, by serendipity, plumes appear as darker regions in the particle images, due to the entrainment of tracer-depleted material from within and above the bottom boundary layer (Fig. \ref{fig:raw}a). When the dark material occupies the boundary layer and a region above it before plume initiation, the dark plume stem width provides an upper bound on the plume stem width. A tighter upper bound comes from the geometry of the dark plume root, which is wider than the stem above it: the height of the sharp turning point between root and stem (star in Fig. \ref{fig:raw}a) approximates the plume stem thickness if the plume material is well outlined \cite{Cagney2015,Cagney2016a,Cagney2016b}. Numerical values are reported in Section \ref{sec:thickness_rslt}.

\subsubsection{From optical distortion}
The same optical distortion that we filter out in Section \ref{sec:velocity_post_process} also encodes information about the plumes that cause it, and can be turned to our advantage as a constraint on the plume stem width \cite{bao2024self}.

Specifically, the refractive index of our fluid decreases as temperature increases (cf. Appendix \ref{app:fluid}), so a plume between the illuminated plane and the cameras casts a ``plume shadow" in the image (Fig. \ref{fig:raw}a); an in-plane plume has negligible influence on the PIV image \cite{Davaille2011}. Using enhanced ray-tracing simulations improved from our previous effort \cite{bao2024self} to account for more realistic experimental conditions (cf. \ref{app:distortion}), we can turn the distortion to our advantage to help us define the lower bound of the plume stem width (cf. section \ref{sec:head_rslt}). The size of the plume head can also be estimated from the optical distortion (cf. section \ref{sec:head_rslt}).

\section{Example Results: Convective flow and plume clusters}
\label{sec:rslt1}
\subsection{Velocity}
\label{sec:velocity}

\begin{figure*}
  \centerline{\includegraphics[width=\linewidth]{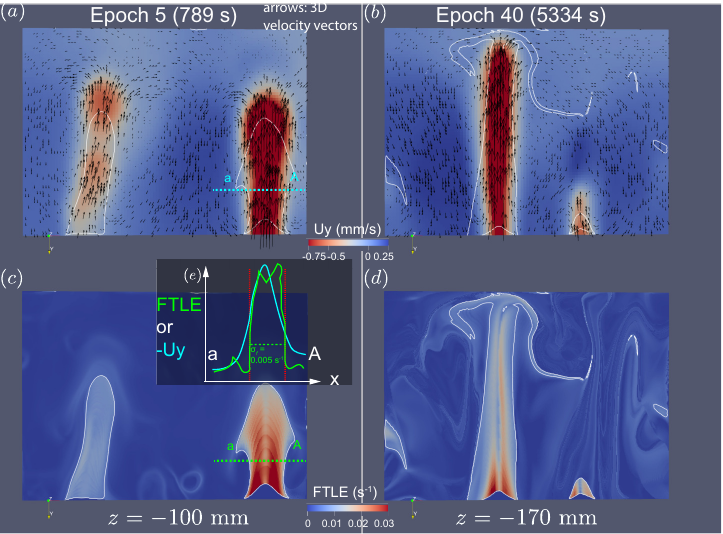}}
  \caption{Cross section of the synchronized vertical velocity $U_y$ in (\textit{a}) and (\textit{b}), with corresponding backward FTLE field $\sigma_f$ in (\textit{c}) and (\textit{d}). The plane is at Epoch 5 (789 s), $z=-100$ mm in (\textit{a}) as well as (\textit{c}), and at Epoch 40 (5334 s), $z=-170$ mm (central plane) in (\textit{b}) and (\textit{d}), respectively. The arrows in (\textit{a}) and (\textit{b}) show the direction and amplitude of the 3D velocity field. In the inset (\textit{e}), $U_y$ (cyan line) and FTLE (green line) along the cross section aA in (\textit{a}) and (\textit{c}) are scaled to the same amplitude and plotted, where red dashed lines indicate the plume boundary from the FTLE profile, determined using the universal threshold $\sigma_f=0.005 \mathrm{s}^{-1}$ (green dashed line). The white contours show $\sigma_f=0.005 \mathrm{s}^{-1}$. FTLE ridges are widened along the plume stem due to limited temporal resolution (cf. \ref{app:FTLE_width}). The spatial extent of the results corresponds to the SSPIV region outlined by the cyan boxes in Fig. \ref{fig:raw}a.}
\label{fig:FTLE_section}
\end{figure*}

\begin{figure}
  \centerline{\includegraphics[width=\linewidth]{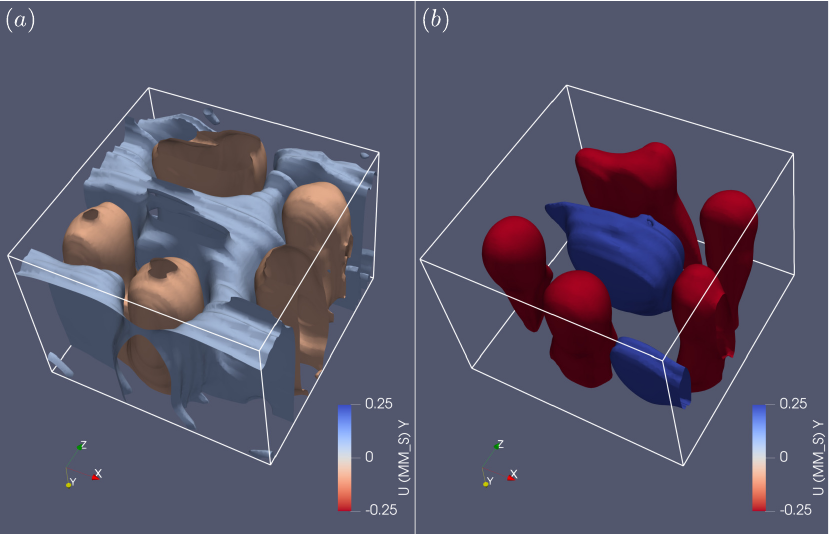}}
  \caption{3D vertical velocity $U_y$ (in mm/s, negative (red) means upwelling) at Epoch 6 (917 s) for the full image volume). Contours are $U_y=$
  (\textit{a})  $\pm$0.075 mm/s and (\textit{b}) $\pm$0.25mm/s, respectively.}
\label{fig:Uy_early}
\end{figure}

\begin{figure}
  \centerline{\includegraphics[width=\linewidth]{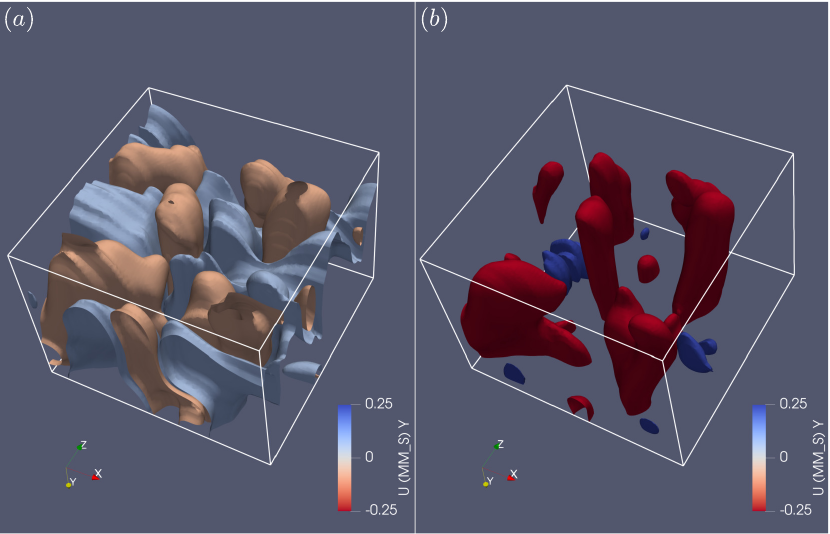}}
  \caption{3D vertical velocity $U_y$ (in mm/s) at Epoch 40 (5334 s). Colors and contours as Fig \ref{fig:Uy_early}.}
\label{fig:Uy}
\end{figure}

The velocity field of the experiment is dominated by the vertical component $U_y$. The 2D cross sections of $U_y$ are plotted for an early (Epoch 6 in Fig. \ref{fig:FTLE_section}a) and a late epoch (Epoch 40 in Fig. \ref{fig:FTLE_section}b). The 3D contours of $U_y$ at the same epochs are plotted in Fig. \ref{fig:Uy_early}, \ref{fig:Uy}, respectively.

We find the PIV domain is quiescent from Epoch 0 to Epoch 2, while upwelling circular domes ("velocity dome") are first observed at the bottom of the tank in Epoch 3 (534 s). They rise to develop into finger-like domes in the velocity space (Fig. \ref{fig:Uy_early}). 

During the ascent of the first batch of 6 plumes, the large-scale downwelling ($U_y$  $\sim$ 0.1 mm/s) occupies the center of the tank, as well as about half of the area near the wall; while the plumes take a ring-like distribution, i.e., they are all located about 135 mm from the center of the tank (Fig. \ref{fig:Uy_early}). As the flow develops, we find, by looking at the velocity contours, more plumes (twice or more as many) spanning the tank at different distances from the center. Those plumes, later in their evolution, are characterized by thinner velocity contour domes, compared to initial plumes (Fig. \ref{fig:Uy}).

Throughout the experiment, we do not find any stable large-scale convection cells or planforms. The region of downwelling can be eroded by new and ongoing plumes, while downwellings can push and deflect the plumes (Fig. \ref{fig:Uy}). 

\subsection{Plume counting and spacing}
\label{sec:spacing}
With more plumes occupying the tank later in time (e.g., Epoch 40 in Fig. \ref{fig:Uy}) compared with the beginning of the experiment (Epoch 6 in Fig. \ref{fig:Uy_early}), the plume spacing is expected to be smaller, which is not easy to measure directly from the velocity field. Thanks to our complete methodology, we can identify the plume velocity centers and their spatial configuration network as a minimum spanning tree (cf. section \ref{sec:count_plume}), with the whole spectrum of plume spacing, i.e., from minimum to maximum for each snapshot.

\begin{figure*}
  \centerline{\includegraphics[width=\linewidth]{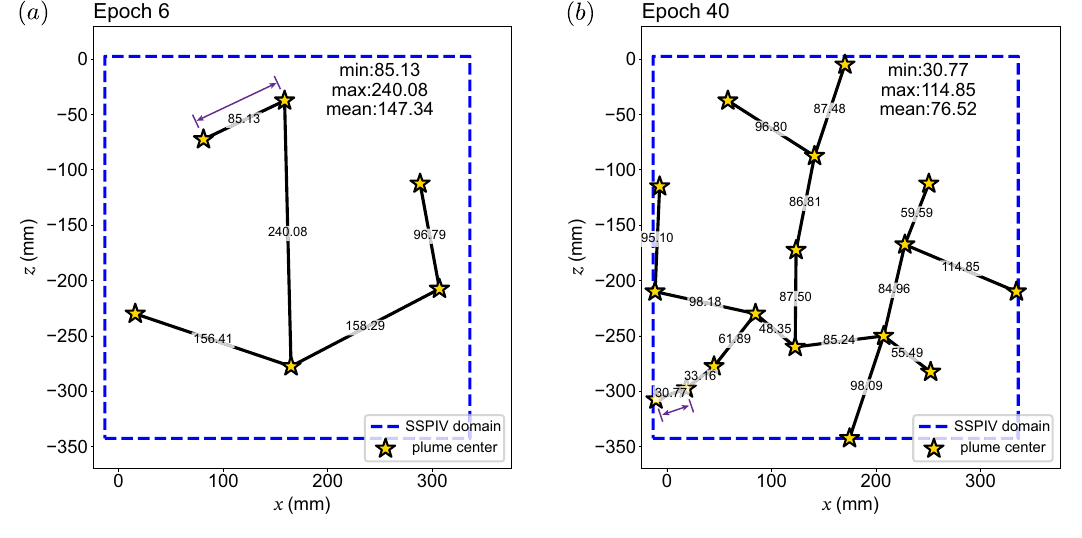}}
  \caption{Minimum spanning tree for (\textit{a}) Epoch 6 (917 s)  and (\textit{b}) Epoch 40 (5334 s). The plume velocity centers (star) are connected in the tree with the black solid lines, with the spacing values shown in mm. The minimum, maximum and mean spacing are shown on the top right of each panel. The purple arrow indicates the pair of plume with the minimum spacing. The blue dashed box shows the coverage of our SSPIV result with respect to the entire tank (the thin solid outline of each panel) in $x$-$z$ direction.}
\label{fig:tree}
\end{figure*}

As an example, Fig. \ref{fig:tree} shows the constructed minimum spanning tree for Epochs 6 and 40. The number of plumes as detected velocity centers increases from 6 (Epoch 6) to 17 (Epoch 40). As a result, the mean plume spacing in Epoch 6 (147.34 mm) is almost twice that of Epoch 40 (76.52 mm), while its minimum spacing (85.13 mm) is close to three times that of Epoch 40 (30.77 mm).

\subsection{Lagrangian analysis of the plume material and plume clusters}
\label{sec:FTLE}
Following the methodology of Sections \ref{sec:lagrangian_method} and \ref{sec:cluster_method}, we now apply the Lagrangian and cluster analyses to our experimental velocity field. The FTLE ridge (high $\sigma_f$) has been previously shown to have sharp edges, when computed from a velocity field with sufficient temporal resolution. The ridges can then be used to objectively track plume materials \cite{Newsome2011,Cagney2015}. In this section, we examine the sharpness of the FTLE ridges in our experiment, when the temporal resolution is not ideal. We also present the first, to our knowledge, attempt at computing 3D FTLE field with multiple plumes, and isolated plume clusters.

\subsubsection{FTLE ridges and the plume structure}
\begin{figure*}
  \centerline{\includegraphics[width=\linewidth]{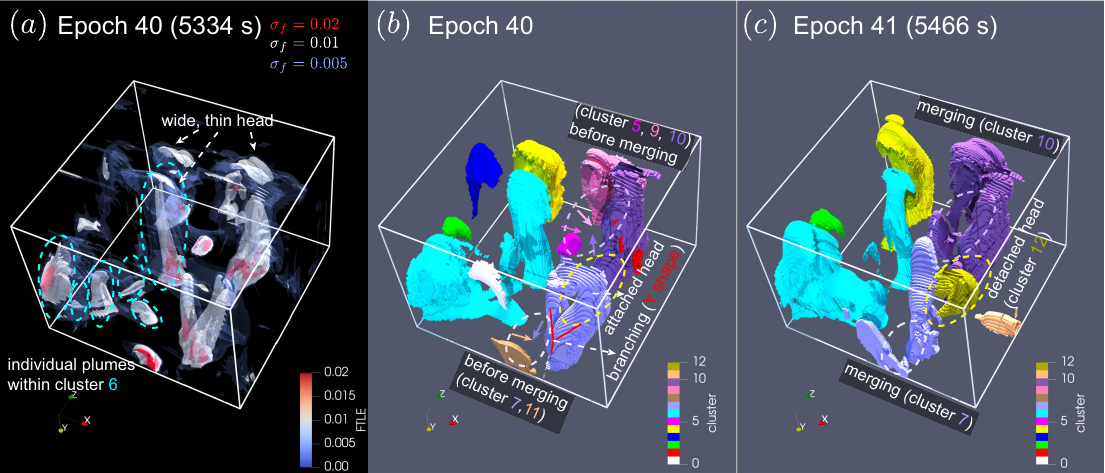}}
  \caption{Comparative view of the 3D FTLE and plume clusters at Epoch 40 (5334 s) and Epoch 41 (5466 s). (\textit{a}) Contours of backward FTLE field $\sigma_f=$ 0.02 (red), 0.01 (white) and 0.005 s$^{-1}$ (blue) respectively, with unit s$^{-1}$, at Epoch 40. Some isolated plumes within one cluster (cluster 6 in \textit{b}) are outlined with cyan ellipses. Arrows annotate the wide, thin head of one plume. (\textit{b}) Plume clusters at Epoch 40. (\textit{c}) Plume clusters at Epoch 41. Cluster numbers are modified to become IDs (identifiers) to be consistent across panels. Some examples of dynamical interactions are highlighted with arrows, dashed ellipses, and labels carrying the relevant cluster IDs in the corresponding cluster colors, like merging (\textit{b}, \textit{c}), attached head branching with a Y shape (\textit{b}), and detached head (\textit{c}).}
\label{fig:FTLE}
\end{figure*}

Examples of 2D backward FTLE fields are shown in Fig. \ref{fig:FTLE_section}c, d. The FTLE ridges, corresponding to the plume material \cite{Cagney2015}, do appear at the same locations of the plumes in the velocity field (Fig. \ref{fig:FTLE_section}a, b). The velocity and FTLE profiles along a representative cross section aA in Fig. \ref{fig:FTLE_section}a, c are shown in Fig. \ref{fig:FTLE_section}e. It is evident that the FTLE ridge has much sharper edges (higher gradient amplitude) compared to the velocity field with varying smooth halos in Fig. \ref{fig:Uy} for example.
Unlike determining the plume boundary using a velocity threshold \cite<or temperature/tracer height fields, >{Cagney2015}, the present boundary is much less sensitive to the FTLE threshold. Therefore, it is possible to objectively isolate a given plume with $\sigma_f$, using, for example, the local minimum value next to the FTLE ridge. To correctly identify the material boundary of a plume, a metric like $\sigma_f$ is necessary, as it reflects the accumulated effects of the flow history, while velocity and its spatial derivative (strain rate) only contain instantaneous information \cite{Davaille2011}. For the same reason, the morphology (shape) of plumes different between velocity and FTLE fields. A 3D backward FTLE snapshot is shown in Fig. \ref{fig:FTLE}a. As one of the best resolved plumes, the central plume in Fig. \ref{fig:FTLE}a nicely highlights the morphology and structure of a plume, including a large thin head, a narrow conduit, and a laterally expanded root at the bottom with the largest $\sigma_f$. Such structure cannot be revealed by the instantaneous velocity snapshot (Fig. \ref{fig:Uy}). The apparent proximity between the FTLE ridge and the half-peak velocity width in Fig. \ref{fig:FTLE_section}c is itself an artifact of the FTLE-ridge widening discussed in Appendix \ref{app:FTLE_width}

Importantly for comparisons to geophysical and geological observations, faster, stronger plumes with larger buoyancy flux have higher $\sigma_f$ \cite{Lin2006b,Cagney2015} (Fig. \ref{fig:FTLE_section}). This means ideally, a series of $\sigma_f$ should be determined for each of the plumes. Meanwhile, the low $\sigma_f$ pattern in the background ($\sigma_f<0.005$ s$^{-1}$, Fig. \ref{fig:FTLE_section}c, d) is the result of the stirring and folding of the ambient material by the large-scale flow and can be effectively removed with a high FTLE threshold to isolate the plume material. However, due to the limited scanning speed (i.e., 130 s per epoch) and temporal resolution of velocity measurements, we find the FTLE ridges are wider than they should be as a result of the velocity interpolation in time during post-processing (cf. \ref{app:FTLE_width}). This widening of FTLE ridges is more severe for faster plumes.

Consequently, we do not try to interpret the width of the plumes with a series of $\sigma_f$ thresholds, instead we use $\sigma_f=0.005$ s$^{-1}$, a small threshold to identify as many plumes as possible and filter out non-plume background structures. For the same reason, we also focus on more robust results, like the statistics of the plume centers (cf. section \ref{sec:spacing}), and the dynamical interaction among clusters of plumes. Alternative methods based on the raw images are available to estimate the plume width (cf. section \ref{sec:thickness_rslt}). 

For completeness, we discuss the widening and other artifacts of our FTLE ridges due to imperfect velocity measurements in Appendix \ref{app:FTLE_width}.


\subsubsection{Plume clusters}
\label{sec:cluster}


Cluster analysis (cf. \ref{sec:cluster_method}) allows us to isolate the FTLE ridges belonging to different groups of connected plumes. As a representative snapshot,
Figure \ref{fig:FTLE}b shows the processed clusters from the raw FTLE field in Figure \ref{fig:FTLE}a. Each plume cluster consists of one to several plumes with various morphologies (Figure \ref{fig:FTLE}b). The separation of the plume material into clusters enables continuous tracking and monitoring of the interplay of different plumes.

As an example, Fig. \ref{fig:FTLE}b,c show 2 consecutive epochs from Epoch 40 to 41, with cluster numbers manually adjusted to be consistent across the 2 epochs. By inspecting the evolution of plume clusters, we find very rich dynamical behavior. For instance, cluster 5,9 and 10 can get closer and merge as a big cluster, similarly for cluster 7, 8, in Epoch 40 (the arrows and white dashed ovals in Fig. \ref{fig:FTLE}b); two plume stems can be connected to the same root as two branches (cluster 7, the red dashed line in Fig. \ref{fig:FTLE}b); the head of a plume can detach from the rest of the plume (cluster 12, the dashed oval in Fig. \ref{fig:FTLE}c). We discuss these results in detail in our companion paper \cite{bao2025evolution}


\subsection{Plume stem width}
\label{sec:thickness_rslt}
Following the two image-based methods introduced in Section \ref{sec:thickness_method}, we now report the numerical bounds on the plume stem width for our experiment. As our FTLE ridges appear wider near the plume stem (cf.  \ref{app:FTLE_width}), we rely on alternative constraints from the particle images themselves to determine the plume stem width (cf. section \ref{sec:SSPIV_process}).

\begin{figure*}
  \centerline{\includegraphics[width=\linewidth]{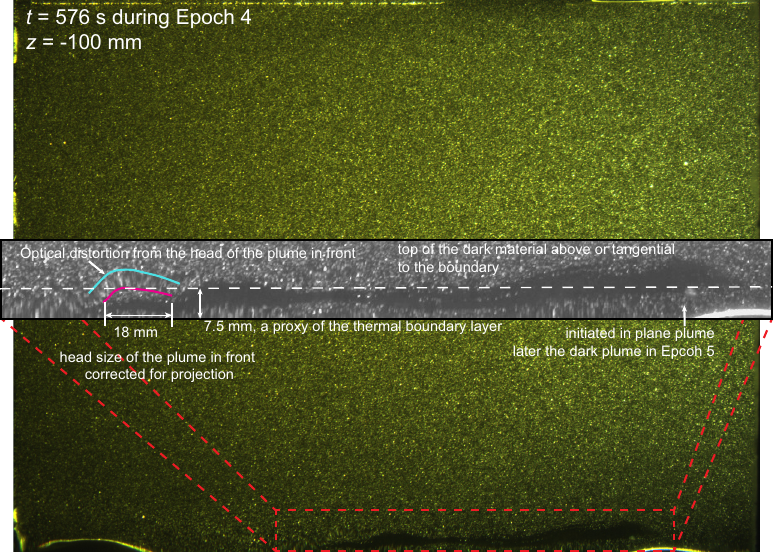}}
  \caption{Particle image from Camera 1 at 576 s during Epoch 4 at $z = -$100 mm. The region around the dark material, later entrained in the dark plume in the same plane in Epoch 5 is zoomed in and shown in gray scale with higher brightness. The upper and lower edges of the optical distortion from the head of the plume in front are outlined in green and pink dashed lines. The height of top of the lower edge is highlighted using the white dashed line, which can be used as a proxy for the TBL thickness (7.5 mm). The top of most of the dark material is above or tangential to the white dashed line. The dark plume in Epoch 5 initiated here, at the right end of the zoomed-in region. The head size of this just initiated plume is 18 mm.}
\label{fig:dark_area_TBL}
\end{figure*}

The dark plume in Fig. \ref{fig:raw}a has a stem width of 8.8 mm, likely an upper bound as it represents entrainment of material both within and above the bottom boundary layer. This extent of the dark material beyond the TBL is supported by the raw image of the same plane one epoch before Fig. \ref{fig:raw}a ($t=576$ s), shown in Fig. \ref{fig:dark_area_TBL}. The top of the dark material was mostly above or tangential to the top boundary of TBL before being entrained into the dark plume in Fig. \ref{fig:raw}a. In terms of the thickness of the TBL $\delta$, a proxy was found as the the height of the lower edge of a plume head just initiated ($\delta\simeq 7.5$ mm), indicated by the plume head optical distortion.

The turning point height of the dark plume stem (detailed view in Fig. \ref{fig:raw}a), yields a 6.8 mm stem width. This is still not a lower bound because of the entrained ambient material. As the second method provides a smaller value, we now have a tight upper bound on the plume width (6.8 - 8.8 mm). Meanwhile, the width of the stem obtained from the optical distortion caused by the plume shadows is $\sim$5 mm across different plumes (Fig. \ref{fig:raw}b). This provides a lower bound according to our new ray tracing simulation (cf. Appendix \ref{app:distortion}). 

\begin{figure*}
  \centerline{\includegraphics[width=\linewidth]{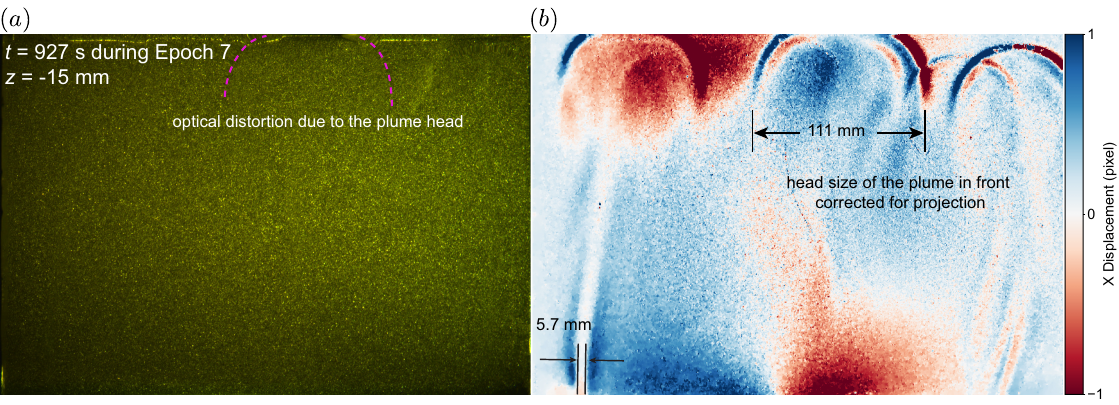}}
  \caption{A particle image (\textit{a}) and corresponding in plane inter-frame displacement (\textit{b}) similar to Fig. \ref{fig:raw} but at 927 s during Epoch 7 at $z=-15$ mm. The plume shadow appears larger than the actual plumes in front, which are closer to the camera than the light plane. The width of the plume head at the center is 111 mm with the scale correction considering the distance difference to the camera.}
\label{fig:raw_head}
\end{figure*}

\begin{figure*}
  \centerline{\includegraphics[width=\linewidth]{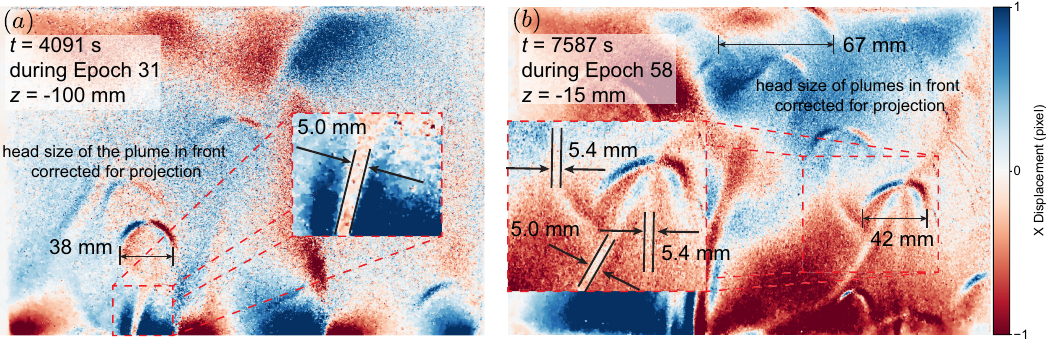}}
 \caption{In-plane lateral inter-frame displacement later in the experiment. (a) 4091 s during Epoch 31 at $z = -$100 mm. (b) 7587 s during Epoch 58 at $z = -$15 mm. The width of the shadow for the stem is still around 5 mm. The scale for the plume heads is corrected considering the projection from the plumes in front to the shadow in the illuminated plane. The typical mid-tank height plume head size is around 40 mm in the latter stages of the experiment, while the head size near the surface is around 65 mm.}
\label{fig:plume_stem_late}
\end{figure*}

Although such distinct dark appearances are not available for most of the plumes in the experimental raw images, we have tracked more plume stem optical distortion (or plume stem shadow) thickness in addition to Fig. \ref{fig:raw}a. We find that its thickness increases slightly from 4.8 mm to 5.7 mm for the same plume in the first batch of plumes (Fig. \ref{fig:raw_head}). But in the middle ($t=4091$ s) and near the end ($t=7587$ s) of the experiment, the plume stem shadow thickness remains around 5 mm (5 to 5.4, Fig. \ref{fig:plume_stem_late}), similar to that in Fig. \ref{fig:raw}a (4.8 to 5.3).

Taken together, the plume thickness $\delta_\mathrm{m}$ appears to be approximately constant throughout the experiment, even though the temperature difference and the heat flux across the bottom TBL should decrease over time as the bulk of the fluid is being heated.

We suggest the typical plume width defined by material boundaries in our experiment is $\delta_\mathrm{m}=6\pm1$ mm based on the dark plume geometry and optical distortion. This is comparable with our determined TBL thickness $\delta\simeq 7.5$ mm. This is also consistent with the material-boundary-defined stem thickness of 6.3 mm reported for a single thermal plume experiment with similar $Ra$ and syrup \cite{Cagney2015}. When scaled to the Earth \cite{bao2025evolution} the plume stem thickness is $\sim65$ km, which is up to 70$\%$ thinner than previous estimates based on the material plume boundary \cite<e.g., the dye-visualized injected-plume experiments in >{griffiths1991dynamics}. In our companion paper \citeA{bao2025evolution}, we further explore the evolution of the TBL thickness $\delta$ with progressively warmer fluid interior from a theoretical perspective.

\subsection{Plume head size}
\label{sec:head_rslt}
The outer edge of a thermal plume head is sharp \cite{Newsome2011,Davaille2011}, i.e., the temperature gradient is high enough to trigger obvious optical distortion in the shadow-affected PIV region \cite{bao2024self}. Unlike the plume stem (cf. Appendix \ref{app:distortion}), the high gradient zone is much thinner than the diameter of the plume head. It is therefore possible to directly infer the plume head size from the raw image and/or PIV result, with just the corrections related to the position of the illuminated plane, the location of the plume head, and the camera. 

We find the width of the plume heads can grow up to about 110 mm before they hit the surface and start to spread underneath the top TBL (Fig. \ref{fig:raw_head}). Note the width can vary significantly for different plumes in space and time during their ascent. Understanding the evolution and size distribution of plume heads requires carefully mapping the plume shadow back to the plumes in front of the illuminated plane, to get the right position and scale, and not all plumes will have clear optical distortion. Nevertheless, we have extracted the head size under representative scenarios. For example, the typical mid-tank plume head size for the first batch of plumes is $\sim60$ mm (Fig. \ref{fig:raw}b), and $\sim40$ in the middle and near the end of the experiment (Fig. \ref{fig:plume_stem_late}), while the size near the surface could be $\sim80$ mm towards the end (Fig. \ref{fig:plume_stem_late}b). The morphology of plume heads can also change with inter-plume and large-scale flow interactions (e.g., central red box in Fig. \ref{fig:raw}b; Fig. \ref{fig:plume_stem_late}). In the companion paper \cite{bao2025evolution}, we will compare the plume head size with theory and we will show the change of the heat flux is mostly reflected in the change of the plume head size and the rising velocity, instead of the plume stem width.

\section{Discussion}
\label{sec:discussion}

\subsection{Improvements on viscous flow measurements}
Thanks to our improved SSPIV system, we are able to measure the full 4D velocity of thermal viscous convection dominated by plumes in our imaged fluid domain. Unlike dyes \cite{whitehead1975dynamics} and shadowgraphs \cite{shlien1976some} which provide access to a projected view, or planar PIV \cite{Davaille2005} which offers in-plane velocity on a given plane, we can directly visualize the evolution of the fluid flow from any angle in the test section. As a result, we can discover the planform of convection, locate individual rising plumes, analyze their temporal and dynamical evolution, and perform a full quantitative analysis at various spatial and temporal scales. Such detailed measurements and analysis methodology help release the power of the infinite resolution of laboratory experiments with real-world physics.

For example, the advantage of our methodology can be demonstrated through perhaps the simplest property of the plumes: their locations (or centers). With a projected view from dyes or shadowgraphs, it is extremely challenging to separate and identify each plume in terms of its original location. With a planar PIV measurement, the center of each plume from the 2D velocity is only approximate, as the exact plume centers are usually not within the illuminated plane. The plume centers obtained from dyes \cite{whitehead1975dynamics} or shadowgraphs \cite{Lithgow-Bertelloni2001} will always be a 2D projection of the 3D locations, so the corresponding plume spacing will be underestimated. Sometimes, velocity or temperature (with thermochromic liquid crystals) of a horizontal plane of the fluid can be measured in experiments \cite{Lithgow-Bertelloni2001,Androvandi2011}, which provides access to the plume locations at a given depth. Note that this is also the common practice when calculating plume spacing in numerical viscous convection simulations, namely considering a selected depth \cite{labrosse2002hotspots,zhong2005dynamics,Arnould2020}. However, our detection of velocity centers (cf. Section \ref{sec:count_plume}) is not limited to a given depth of the fluid; it can capture velocity centers at every depth in a 3D volume, automatically accounting for complex plume morphology and evolution. Therefore, our measurements and methodology can provide a more comprehensive and robust configuration of the plume location and structure.

\subsection{Definition of the plume boundary}
To analyze the morphology or the stirring associated with plumes, it is not enough to obtain the velocity field, it is necessary to accurately and objectively define the boundary of plumes. In experimental or numerical simulations of mantle plumes, it is a very common practice to use velocity or temperature to define the boundary of plumes \cite{Davaille2011,Arnould2020}, but this suffers from at least two drawbacks. First, the velocity and temperature contours of the same plume are distinct \cite{Davaille2011,Cagney2015}. Second, the criteria for threshold selection (for velocity or temperature) is subjective. The amplitude of velocity changes smoothly from the surrounding flow to the interior of plumes (Fig. \ref{fig:FTLE_section}a, b). Fig. \ref{fig:Uy_early}, \ref{fig:Uy} clearly show how different contour selection thresholds can change the perceived size, morphology, and connectivity of the plume domes. The instantaneous snapshot of velocity contours does not provide the history-dependent material transport with which to understand the entrainment, stretching, and stirring, crucial for elucidating the plume source, and its relationship to the overall flow, facilitating comparison to geological observations. The Lagrangian analysis in section\ref{sec:lagrangian_method} provides a more objective measure of the plume boundary, as it appears sharp in the FTLE field (Fig. \ref{fig:FTLE_section}). Such a measurement can help reconcile the varying interpretations and plume scalings from different studies and different measurement techniques  \cite<e.g.>{Davaille2011}.

\subsection{Richness of the plume dynamics and the source of plumes}
The example results shown here (e.g., Fig. \ref{fig:FTLE}) are only a minimal sample of the rich plume dynamics that can be explored with this full end-to-end methodology. Lagrangian and cluster analysis not only reveal but quantify a great variety of plume morphologies and interacting dynamics among plumes, like merging, branching, and detachment, which we delve into further in our companion paper \citeA{bao2025evolution}. These behaviors parallel independent geological and geophysical evidence for similar complexity in the mantle, including hotspot motion \cite{konrad2018relative}, tree-like low seismic velocity structures in the deep mantle \cite{Tsekhmistrenko2021}, and possibly seismically imaged dying plumes \cite{Silveira2006}, suggesting the dynamical richness we observe in the laboratory is relevant to the real mantle rather than only an artifact of finite experimental geometry. Such interactions and variable morphologies, some previously seen \cite<e.g.> {scott1986observations,loper1986mantle,olson1986solitary,Moses1993,Manga1993,Kelly1997,Davaille2002,Gonnermann2004,Davaille2005}, outline potential pathways for plume evolution. These are important because the paths and interactions of individual plumes and the surrounding flow, will affect material entrainment and therefore the interpretation of geochemical anomalies at the surface. For example, advection of numerical tracers within the observed 4D velocity, and computed FTLE ridges, will allows us to map a tracer within the plume to its source location, getting us closer to a full surface-to-source mapping from hotspot to plume source. With future improved velocity measurements (e.g., finer temporal resolution and expanded spatial coverage), our methodology has the potential to connect the geochemical variability observed in ocean islands to the possible reservoirs at the bottom of the mantle. The present experiment provides a time-resolved ensemble of plume configurations used here and in the companion paper \cite{bao2025evolution} to quantify plume counts, spacing, recurrence timescales, and evolutionary pathways. We also see similar plume behaviors in experiment with similar conditions \cite{Bao2024PhD} and plan to reproduce each of the different plume behaviors , and a series of controlled experiments with 4D quantified measurements. 

Although our experiment is not intended as a one-to-one model of mantle convection, it isolates a dynamically relevant end-member for mantle plumes: inertia-free, high-$Ra$, high-$Pr$, temperature-dependent viscous flow with multiple coexisting plumes generated from an unstable basal thermal boundary layer. Several differences from the natural mantle system should be made explicit. First, the basal temperature in our experiment is laterally uniform and held fixed after the initial heating ramp, whereas the CMB evolves as the core cools over geological time. Second, the real CMB is laterally heterogeneous, with structures such as Large Low Shear Velocity Provinces, Ultra-Low Velocity Zones, the post-perovskite phase boundary, and possible chemical interaction with the outer core; these complexities are absent in our uniformly heated tank. Third, our upper boundary is a rigid lid rather than a mobile plate boundary with plate-driven stresses and lateral flow. Finally, the experiment does not include secular mantle cooling, compositional heterogeneity, or long-term chemical evolution. Its value is therefore not as a complete mantle analogue, but as a controlled system in which plume spacing, morphology, and interaction can be quantified under well-defined high-viscosity conditions. In this regime, transient changes in plume morphology arise naturally from plume-plume interaction, boundary-layer variability, and large-scale return flow, suggesting that natural mantle plumes need not remain isolated or morphologically steady, and that geophysical or geochemical observations may sample transient or composite plume states rather than idealized single conduits.

\subsection{Limitations and future directions}
With our current analysis methodology, we can robustly extract the center locations and therefore the number and spacing of plumes. A network of plumes can be constructed, to analyze them directly in the full 4D space. We identify the various spatial/temporal scales of the changes of the plume centers, as well as inter-plume interactions in a companion paper \citeA{bao2025evolution}.

The FTLE ridges in this experiment are too wide because of the limited time resolution, and our velocity coverage in space near the tank walls is not as good as in the interior. Apart from improving the SSPIV system, it might be also possible to incorporate the existing PIV velocity data to construct a digital twin of our current viscous convection experiment, with which the unseen interior temperature field can be reconstructed \cite{Bao2024PhD}. Constructing a digital twins, temperature fields, and addressing material entrainment and metrics will be the subject of upcoming papers.

On the other hand, the raw particle images provide independent constraints on the plume stem width from the dark region and optical distortion. The morphology and evolution of the plume shadow, in particular, can be also used to perform a one-to-one comparison with the FTLE ridges after improved velocity measurements or for the digital twin.

\section{Conclusions}
\label{sec:conclusion}
We presented the 4D velocity measurement of a Rayleigh-B\'enard experiment in a very viscous fluid with multiple interacting laminar thermal plumes, using a 3D Scanning Stereoscopic PIV (SSPIV) system. We also carefully mitigated for the effects of optical distortion in PIV due to the presence of hot, buoyant plumes, which may have hindered previous attempts \cite{Laudenbach2001,bao2024self}. We constructed, an end-to-end methodology based on Lagrangian and cluster analysis, K-D Tree, and minimum-spanning tree metrics, and obtained a full 3D FTLE field with multiple plumes. Despite the limited temporal resolution of the velocity data, the FTLE ridges still have sharp edges, and enable more objective determination of the material boundary of the plumes compared with traditional metrics like velocity, temperature, and tracer height contours. Our analysis methodology further isolated plumes from the background using the FTLE ridges, separated them into disconnected plume clusters, located the plume centers, and identified the full spectrum of plume spacing. We also obtained constraints on the plume thickness from the tracers and optical distortion in the particle images as well as the FTLE fields, which show plumes to be much thinner thanprior work that also used plume material boundaries \cite{griffiths1991dynamics}. To demonstrate the power of SSPIV and our analysis methodology we showed example visualizations of the velocity field at different times, as well as FTLE ridges corresponding to individual plumes. The convection planform and the interaction among plumes are easy to see in our results. Our cluster analysis reveals very rich plume dynamics, including but not limited to merging, splitting, branching, and detachment. 

The contribution of this study to fluid dynamics and geophysics lies in the integration and validation of an end-to-end measurement and analysis framework (SSPIV plus LCS plus cluster analysis) applicable across laminar convection and other low Reynolds flows. Hence, while the use of SSPIV or any individual component or technique is not entirely new (although the cluster analysis and minimum spanning tree adopted here have not been used before in plume studies), the assembly of a complete end-to-end methodology --- broadly applicable across disciplines and numerical simulations --- is new and motivated us to present it as a dedicated paper. Overall, our method enables quantitative analysis of the collective plume dynamics and plume evolution across the test section in unprecedented detail. This methodology, equally applicable to numerical simulations, can be used in future improved measurements for Earth and planetary studies, and to track convective features in various laminar flow experiments, like biological locomotion \cite{wilson2009lagrangian}, blood flow \cite{badas2013use} and chemical reactors \cite{llamas2020potential}.
%
%

\section*{Availability Statement}
Jupyter Notebooks for the plume analysis can be found at \citeA{bao_2025_17489520}. The modified version of \texttt{photon} \cite{rajendran2019piv} used in this study to track optical distortion due to plumes is available at \citeA{bao_2025_15128024}.

\section*{Conflict of Interest declaration}
The authors declare there are no conflicts of interest for this manuscript.

\acknowledgments
The project was directly supported by the National Science Foundation under grant EAR-1900633 to C.L-B. C.L-B. was further supported by the Louis B. and Martha B. Slichter Endowed Chair in Geosciences. X.B. was further supported by Harvard Reginald A. Daly Postdoctoral Fellowship.

%
%

\appendix
\section{Properties of the fluid and tank}
\label{app:details}
\subsection{Fluid property measurements}
\label{app:fluid}
\begin{figure}
  \centerline{\includegraphics[width=0.8\linewidth]{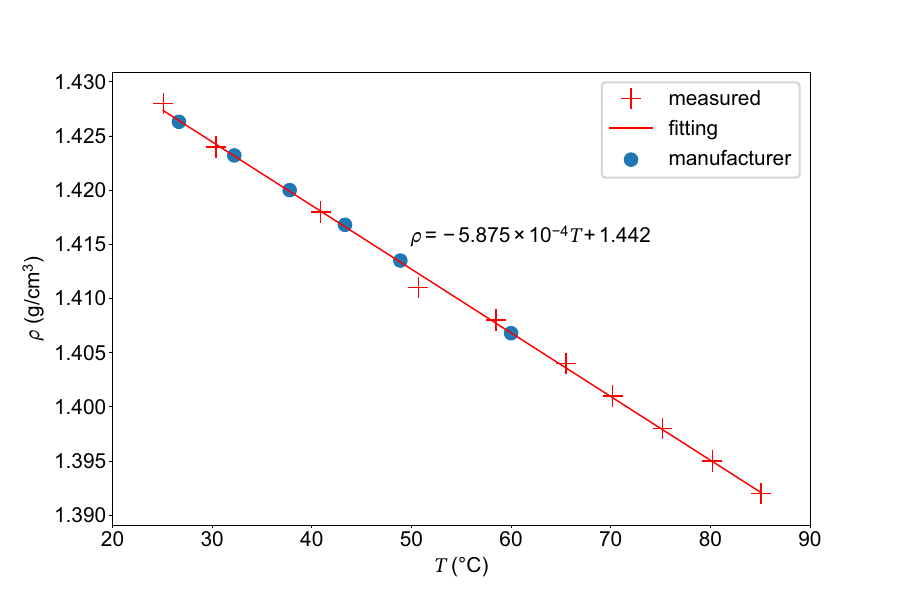}}
  \caption{Density, $\rho$, of Gateway Du-Crose 3 63/43 corn syrup as a function of temperature.}
\label{fig:density}
\end{figure}

The working fluid is the corn syrup Gateway Du-Crose 3. The density $\rho$ 
(in g/cm$^3$) at different temperature (in $^\circ$C) was measured using the Cole-Parmer Specific Gravity Hydrometer Set, results are shown in Fig. \ref{fig:density}. All temperature measurements were made using a K-type thermocouple. The density can be fit by $\rho=-5.875\times 10^{-4}T+1.442$. Our measurement is in remarkable agreement with that from the manufacturer. 

\begin{figure}
  \centerline{\includegraphics[width=0.75\linewidth]{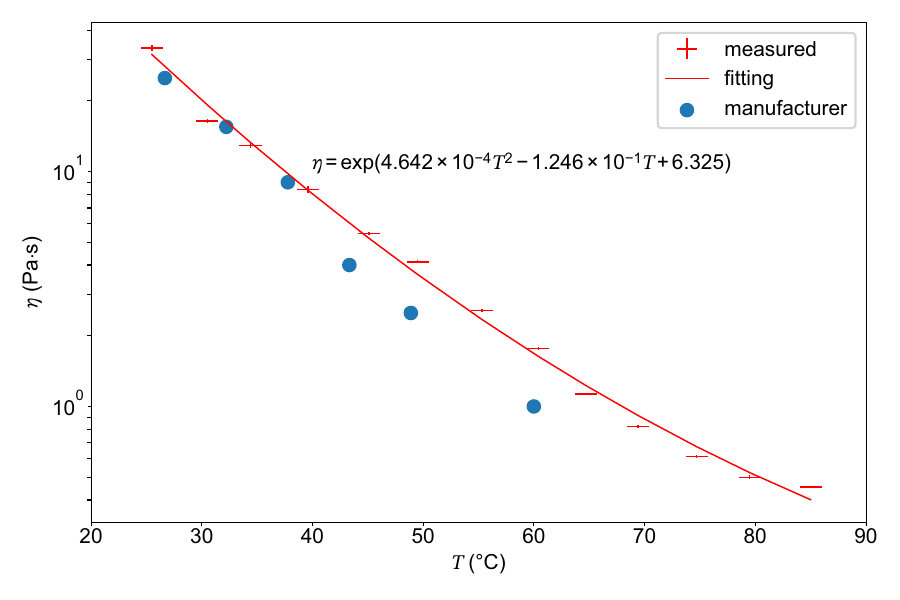}}
  \caption{Viscosity, $\eta$, of Gateway Du-Crose 3 63/43 as a function of temperature.}
\label{fig:viscosity}
\end{figure}

The dynamic viscosity was measured with the Ametek Brookfield DV2TLV viscometer. The viscosity of corn syrup can vary with each batch, and our measurement agrees with manufacturer reference values below 40 $^\circ$C, and is slightly higher above 40 $^\circ$C (Fig. \ref{fig:viscosity}).

\begin{figure}
  \centerline{\includegraphics[width=0.8\linewidth]{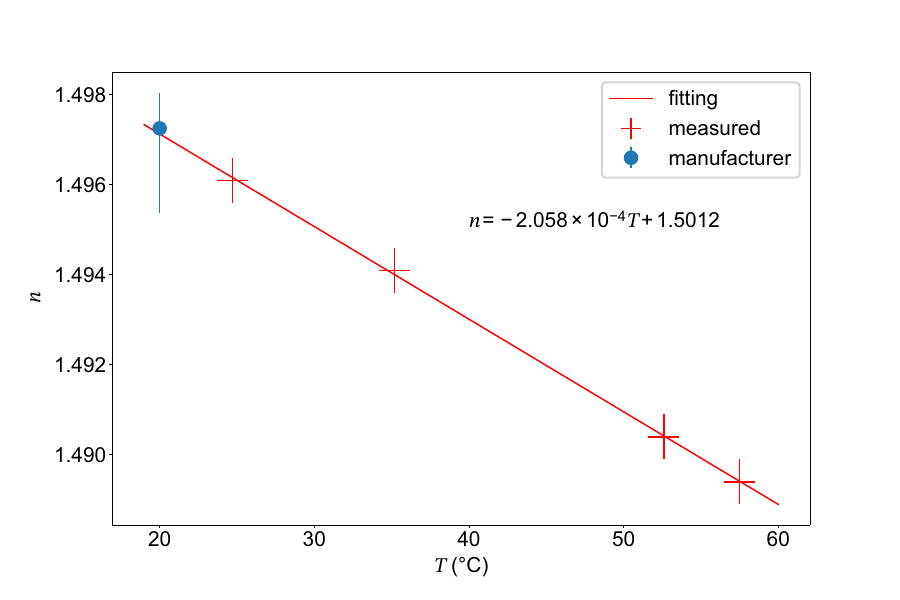}}
 \caption{Index of refraction $n$, of Gateway Du-Crose 3 63/43 corn syrup as a function of temperature.}
\label{fig:IOF}
\end{figure}

The refractive index $n$ as a function of temperature was obtained using a Hanna instruments refractometer HI96800.  Our measured data is fit by $n=-2.058\times10^{-4}T+1.5012$, which matches the reference value from the manufacturer at 20 $^\circ$C (blue dot in Fig. \ref{fig:IOF}). This suggests a solid substance content of $81.7\%$, consistent with the manufacturer data at this temperature ($81\%$ to $82\%$). The properties have been summarized in Table \ref{tab:syrup}. 

To reduce the impact of possible heterogeneities in the syrup, before each experiment, the syrup was first mixed by a Cole-Parmer Compact Digital overhead stirrer, and then basally heated at 80 $^\circ$C for 8 hours, while in the tank, to ensure chemical homogeneity and allow air bubbles to rise. After homogenization and stirring, we let the fluid sit in the tank at room temperature for one week to cool down and expel the air. 

\begin{table}
\caption{Corn syrup properties at 25$^\circ$C. $C_P$ and $k$ are from the manufacturer. Reproduced from \citeA{bao2024self}, with the permission of AIP Publishing.}\label{tab:syrup}
\begin{tabular}{@{\extracolsep\fill}lcr}
\hline \hline
Properties  & Symbol   &   Value  \\[3pt]
\hline
Density & $\rho$ & 1427 $\mathrm{kg/m}^3$ \\
Dynamic viscosity  & $\eta$ & 33.1 Pa$\cdot$s \\
Kinematic viscosity  & $\nu$ & 0.023 m$^2$/s \\
Index of refraction  & $n$ & 1.4961 \\
Thermal expansion coefficient & $\alpha$ & $4.1\times 10^{-4} \mathrm{K}^{-1}$ \\
Thermal conductivity & $k$ & 0.346 W$\cdot$m$^{-1}$K$^{-1}$ \\
Specific heat capacity & $C_P$ & 2300 J/kg$\cdot$K \\
Thermal diffusivity & $\kappa = \frac{k}{\rho C_P}$ &  $1.05\times 10^{-7}$ m$^2$/s \\
Prandtl number & $Pr$ & $2.2\times 10^5$ \\
\hline \hline
\end{tabular}
\end{table}

\subsection{The wall of the test section}
\label{app:wall}
The acrylic wall (10 mm thick ACRYLITE GP acrylic sheet) has a thermal diffusivity of $10^{-7}$ m$^2$/s and a refractive index $\sim$ 1.49, both nearly identical to that of the working fluid described above. The tank provided effectively insulating walls and a no-slip boundary condition on all boundaries. The acrylic grade was high, allowing $\>$92\% visible light (400-700 nm) transmission, enabling photographic access and optimal brightness of the image planes for PIV.

\section{Calibration of the SSPIV system}
\label{app:calibration}
\subsection{Positioning Tests}
\label{app:pos_calibration}
Confirming the reliability and potential error of the carriages returning to the same position on the linear slides is necessary to ensure accurate SSPIV measurements. Following \citeA{Pears2015}, 5 tests were performed three times using an axial dial indicator DITR-0105. The average results show that in general the error is on the order of microns, which is $\sim 1$‰ of the 5 mm step interval (Table \ref{tab:positioning}). This is very precise, especially compared with the thickness of the light sheet ($\sim 5$ mm). The synchronous movement of the cameras and the light sheet is controlled by the computer with microsecond-level accuracy.

\begin{table}
\begin{center}
    \begin{tabular}{p{2.5cm}p{6cm}c}
    \hline \hline
    Test  & Description (To test if …)  & Precision ($\mathrm{\mu m}$)\\[0.5ex]
    \hline\\[-1.5ex]
    Home   & Carriages can return to home position with precision.    & 0.6\\
    Accuracy &  Carriages can move forward (5 mm) with precision. & 10\\
    Backlash & Forward and backward steps are both accurate. & 8.33\\
    Long PHO & Carriage can rebound to the correct PHO (200 mm) after reaching home position by clicking home command in SPIVET-Control. & 4.75\\
    Step Repeatability & Carriages can move forward 10 steps with precision. & 4\\
    \hline \hline
    \end{tabular}
\end{center}
\caption{Positioning test results for the scanning system. Results are the average over 5 repeated tests.}
\label{tab:positioning}
\end{table}

\subsection{Photogrammetric Calibration}
\label{app:photo_calibration}
\begin{figure}
  \centerline{\includegraphics[width=\linewidth]{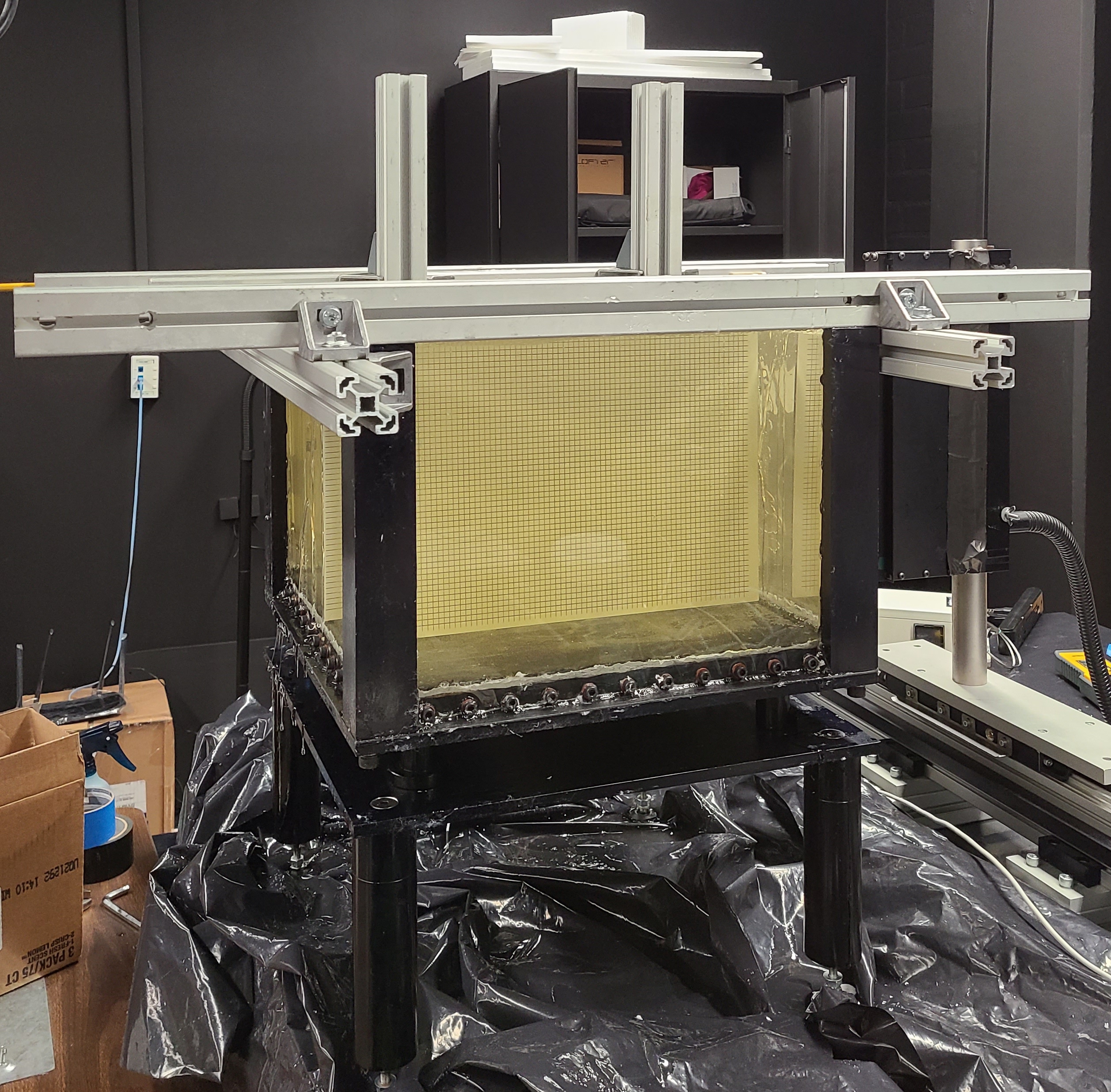}}
    \caption{The setup of the calibration target (5 mm grid) in the tank with syrup.}
    \label{fig:caltarget}
\end{figure}

SSPIV photogrammetric calibration is necessary to map the image coordinates of the camera CCD sensor to the coordinates in the laboratory. This was done with a 5 mm-gridded calibration target submerged into the working fluid, parallel to the light sheet and perpendicular to the cameras. A rigid structure is available to allow the target to be fixed as needed and move with the light sheet (Fig. \ref{fig:caltarget}). The cameras were set to image a common region, and the focal length were chosen to allow the largest field of view possible. The gridded target was placed at the center plane, 2 in front of it, and 2 behind it, each with about 5 mm interval. At each position, photos were captured, and precise displacement of the target was recorded.

The photos were then edited to enhance the brightness, and remove obstructions to grid intersections. We used SPIVET-UCLA to extract intersections, and built a modified pin-hole model \cite{tsai1987versatile} to map 2-D pixels in the image to 3-D point coordinates in the lab, so we could dewarp the image, i.e., project the raw image to mid-plane of the light sheet. Note that a smaller step interval of the cameras ($\sim$3.4 mm) than that of the light sheet (5 mm) was used during SPIV scanning. The ratio of the light step interval over the camera step interval is very close to the refractive index of the working fluid. This ensured the size of illuminated planes to be constant in the images, and the photogrammetric calibration at the central plane could be used at all plane positions. The final photogrammetric error is $1.96\pm1.21$ pixels and $1.76\pm1.05$ pixels for camera 0 and 1, respectively.

\section{Taking Advantage of Optical Distortion}
\label{app:distortion}
The hotter temperature within the stem of the $p_\mathrm{if}$ reduces the density ($\rho$) and refractive index ($n$)  (Fig. \ref{fig:density},\ref{fig:IOF}) towards the center of the stem. The change in $\rho$ and $n$ lead to optical distortion in the particle image and the PIV displacement obtained with optical flow (Fig. \ref{fig:raw}). In this section, we will use the optical distortion to help constrain the diameter of the plume stem, which is a material or entrainment boundary.

\subsection{Theoretical analysis}

\citeA{Laudenbach2001} suggest that the temperature (hence $\rho$ and $n$) along the stem of a thermal plume roughly follows a Gaussian distribution at different radii at a given stem height. The deflection angle as the light ray travels through a density gradient zone is proportional to $\nabla n$ integrated along the ray trajectory. \citeA{elsinga2005evaluation,Laudenbach2001} further showed that 
maximum deflection occurs when the light ray passes 1$\sigma$ of the stem center. Therefore, the maximum distortion we see in the particle image near the plume stem is the distance 1$\sigma$ away from the stem center. Ideally to use the observed distortion we would need 1) the illuminated plane to just touch the edge of the stem of the  $p_\mathrm{if}$, and 2) the camera to be at the same $x$ location as the $p_\mathrm{if}$. 

In our experiments, the real configuration of the illuminated plane, $p_\mathrm{if}$, and cameras does not follow the ideal case, but the large distance along $z$ between the cameras and the $p_\mathrm{if}$ makes their $x$ offset negligible. However, it is hard to locate the largest distortion in the particle image. Instead, when the integrated $\nabla n$ is larger than a certain threshold, elongation of tracers occurs. We can use this elongation to our advantage. Three bands can be usually observed around the stem of the $p_\mathrm{if}$: a central band with small and subtle optical distortion, and two distortion bands on the sides with tracer elongation. The distortion bands cover the region within 1$\sigma$ of the stem center on both sides of the plume stem. If the illuminated plane just touches the edge of the stem of the $p_\mathrm{if}$, then the distance between the two distortion bands, i.e., the width of the central band, should be smaller than 2$\sigma$.  The width can be measured at the pixel level using optical flow (Fig. \ref{fig:raw}b). However, when the illuminated plane is far from the $p_\mathrm{if}$, the plume shadow and hence the central band, will appear larger in the particle image. The width of the central band with respect to the 2$\sigma$ is therefore uncertain unless more quantitative evaluation is available.

\subsection{Numerical simulations}
\begin{figure*}
  \centerline{\includegraphics[width=\linewidth]{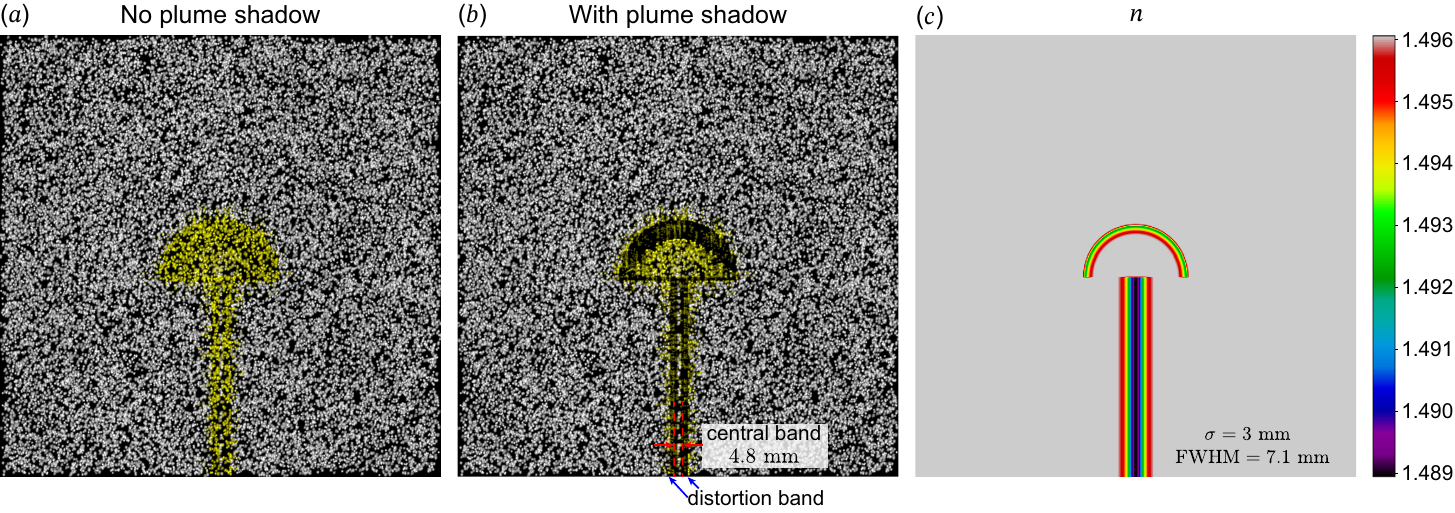}}
  \caption{Synthetic particle images and setup of the refractive index field. (a) Particle image with no plume shadow. (b) Particle image with the plume shadow cast by the structure in (c). Differences between (a) and (b) are highlighted in yellow. The stem part of the plume shadow contains a central band (4.8 mm width) and two distortion bands. (c) Cross-section of the refractive index field containing the plume-like structure. The anomaly in the stem part is a Gaussian, with $\sigma=3$ mm and FWHM = 7.1 mm (Full-Width-Half-Maximum).}
\label{fig:ray-tracing}
\end{figure*}

To better constrain the width of the central band, we perform numerical simulations to get synthetic particle images that include a plume shadow, using \texttt{photon} \cite{rajendran2019piv}, a Python-CUDA program powered by fourth-order Runge-Kutta ray-tracing schemes optimized for GPUs. We consider a smaller subset of the illuminated plane (22.5 cm$\times$22.5 cm), and a single camera facing the center of the subset, while the rest of the configuration exactly follows our experiment. Our modeling procedure largely follows that in \citeA{bao2024self}, with the key difference that the 3D refractive index field (with 0.6 mm grid resolution) now includes both the syrup and the air (tank wall has essentially the same $n$ as the syrup), instead of syrup only with some approximations. The initiation of light ray angles has been optimized in this work to let tracers cast enough light rays (over 1 billion) to the camera even after the $n$ jump at the tank-air interface. The $p_\mathrm{if}$ is modeled as a plume-like structure, including a cylindrical stem and half a spherical shell head (cross section shown in Fig. \ref{fig:ray-tracing}c). The cylinder has a Gaussian $n$ negative anomaly with a peak value of $pk_\mathrm{stem}$ and a standard deviation of $\sigma$, while the spherical shell has an average radius of $R_\mathrm{head}=24$ mm, and a thickness of 6 mm. $n$ decays quadratically from $0.6pk_\mathrm{stem}$ at 1/4 thickness from the top to 0 at the edge. For simplicity, no variation of $n$ along $y$ is imposed.

Fig. \ref{fig:ray-tracing} shows an example of the central plane of the tank been illuminated. In Fig. \ref{fig:ray-tracing}a, no $p_\mathrm{if}$ is present, while in Fig. \ref{fig:ray-tracing}b, a $p_\mathrm{if}$, is added 13 cm in front of the illuminated plane, and placed in the center of the plane along $x$. The  $p_\mathrm{if}$ has a $pk_\mathrm{stem}$ corresponding to a plume temperature, $T=60~^\circ$C, $\sigma=3$ mm, and FWHM (Full-Width-Half-Maximum) = 7.1 mm (Fig. \ref{fig:ray-tracing}c). The two distortion bands and the central band are successfully reproduced (Fig. \ref{fig:ray-tracing}b). The width of the central band is 4.8 mm, similar to those seen in our experiment (Fig. \ref{fig:raw}), and smaller than 2$\sigma=6$ mm and the FWHM.

We also find central band widths consistently smaller than 2$\sigma$ under various conditions within the range of our experiment. For example, with different illuminated plane and $p_\mathrm{if}$ position/distance, different $\sigma$, and different peak temperature (e.g., 50 to 80$~^\circ$C). Even considering the finite resolution in our $n$ mesh, we can conclude that the central band width is smaller than 2$\sigma=6$ in our experiment.

The final step involves converting this lower bound obtained from a thermal definition of the plume stem boundary to the material boundary of our FTLE ridges. Numerical simulations have suggested that the edge of the FTLE ridge of the plume stem roughly corresponds to FWHM for an axisymmetric thermal plume \cite{Lin2006b}. We therefore think it reasonable to use the central band width as a lower bound of the plume stem thickness defined by material entrainment.

\section{The width of FTLE ridges associated with plumes affected by time resolution}
\label{app:FTLE_width}
Comparing the raw image in Fig. \ref{fig:raw}a, and the FTLE field of the same plane (Fig. \ref{fig:FTLE_section}c) at a similar time (704 s versus 789 s), one might notice that the dark plume in the raw image is considerably thinner than the same plume outlined by $\sigma_f$. The thickness of the plume stem from $\sigma_f$ is about 4 to 5 times that of the dark plume in the raw image (i.e., 8.8 mm). This discrepancy is a natural consequence of the finite time resolution of the experiment and the velocity interpolation in time.

During the data acquisition, both the scanning and exposure take a finite amount of time. While each epoch takes about 130 s, the exposure takes less than 0.1 s for each image and $<12.5$ s for one epoch with an inter-frame interval $<1.5$ s. Although our stepper motors can scan as fast as 1 m/s (and scan the full test section in as little as 30 s), it takes time to accelerate and stop the carriers. Very fast scanning introduces significant vibrations given the load of the camera arm (Fig. \ref{fig:lab}d, f) which could be unstable and unsafe. An additional waiting period between motions is needed before exposure to suppress vibrations and acquire clear images. Reducing the 130 s cycle below its present value would therefore require either a lighter camera mount (with all the optical-alignment consequences of redesigning it) or accepting blurred frames and a degraded velocity field. Reducing the number of planes per epoch is the other lever, but degrades the spatial coverage that this multi-plume study requires. For the present experiment we judged the present cycle time 130 s to be the right trade-off. This is enough time for a plume to travel up to 1/3 of the tank height. Therefore, the synchronization process which interpolates the PIV of different planes to the last plane, and the interpolation during passive tracer advection can lead to some artifacts in the velocity field as well as the FTLE field. This is especially true for planes far from the targeted plane for synchronization.

\begin{figure}
  \centerline{\includegraphics[width=\linewidth]{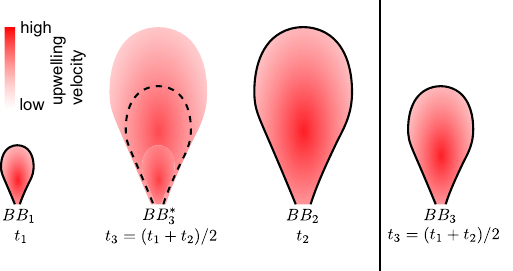}}
  \caption[The artifact from interpolating plume velocity in time]{Cartoon depicting the overestimated plume size arising from interpolating the plume velocity in time. A perfect axisymmetric plume is represented by self-similar oval-like bubbles at times $t_1$ and $t_2$ ($t_2>t_1$), denoted as $BB_1$, $BB_2$, respectively. At $t_3=(t_1+t_2)/2$, the upwelling interpolated from $BB_1$ and $BB_2$ is shown as $BB_3^*$, while the expected ``true" result should be $BB_3$.}
\label{fig:ring_bubble}
\end{figure}

\begin{figure}
  \centerline{\includegraphics[width=\linewidth]{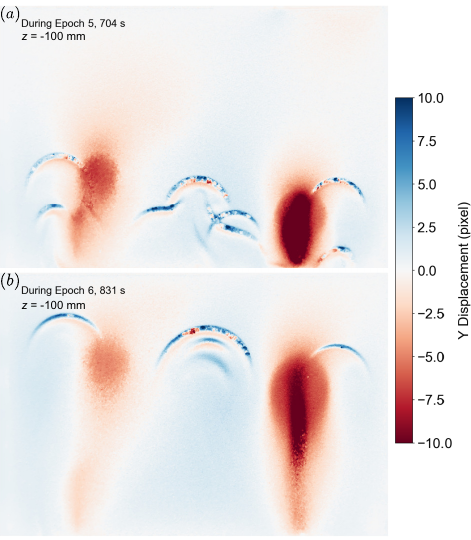}}
  \caption{The raw in-plane Y displacement for $z = -100$ mm. (a) During Epoch 5, $t=704$ s, before the end of Epoch 5; and (b) during Epoch 6, $t=831$ s, after the end of Epoch 5.}
\label{fig:E5E6raw}
\end{figure}

To understand the possible artifacts from the synchronization, consider the $U_y$ of an axisymmetric rising plume. The upwelling region (e.g., Fig. \ref{fig:FTLE_section}a, b) can be characterized by an oval-shape bubble \cite{Cagney2015}, and has been shown to reside mostly in a self-similar ``vortex ring bubble" \cite{Cagney2016a}. Hence, we can approximate the upwelling velocity distribution with an oval-like bubble. As the plume develops, the size and velocity amplitude of this bubble will grow with time (Fig. \ref{fig:ring_bubble}). Consider two snapshots of the velocity field at times $t_1,t_2$ (e.g, $t_2-t_1>100$ s here) for the axisymmetric rising plume, with corresponding bubbles $BB_1, BB_2$, respectively, and no other flow (zero velocity background). The interpolated upwelling bubble $BB_3^*$ one obtains by interpolating between $BB_1, BB_2$ at time ($t_3=(t_1+t_2)/2$) will have the same size (or area) as $BB_2$. This size, however, will be larger than the expected size of the ``true" bubble $BB_3$ at $t_3$, which should be between $BB_1$ and $BB_2$. Multiple local upwelling velocity maxima could arise in $BB_3^*$ (Fig. \ref{fig:ring_bubble}), as seen in the plume on the left in Fig. \ref{fig:FTLE_section}a. Fig. \ref{fig:E5E6raw} shows the raw velocity from optical flow without plume shadow correction is presented for the same plane as Fig. \ref{fig:FTLE_section}a, before and after the synchronization time, i.e., the end of the Epoch 5. The flow field for the two plumes are more complicated than the ring bubble, but largely consistent with the processed, interpolated result in Fig. \ref{fig:FTLE_section}a, especially for the large, circular structure near the top of the left plume, and tilted, stretched and almost disconnected lower part beneath, although the interpolated upwellings occupies extended regions covering those in both raw snapshots, as expected by the simple velocity bubble. The overestimated velocity bubble size causes the FTLE ridge to be much wider than it should be, especially for the plume stem ( Fig. \ref{fig:FTLE_section}c). On the other hand, the size of the plume head from FTLE ridges is not similarly affected.

When a plume is fully developed and the size of the velocity bubble does not change significantly between two epochs, the width of the FTLE ridge is no longer significantly overestimated (e.g., the left plume in Fig. \ref{fig:FTLE_section}b, d, or the central plume in Fig. \ref{fig:FTLE}a).

\begin{figure}
  \centerline{\includegraphics[width=\linewidth]{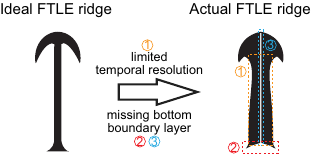}}
  \caption{Artifacts in the FTLE ridge discussed in this study.}
\label{fig:FTLE_artifact}
\end{figure}

In Fig. \ref{fig:FTLE_artifact}, we summarize the artifacts in the plume FTLE ridge, including the larger stem width (artifact 1) due to the insufficient temporal resolution and synchronization, as well as the gap at the root and the center of the stem (artifact 2 and 3) due to the bottom boundary layer being excluded from the SSPIV domain to avoid excessive optical distortion.

The center along the plume stem usually shows lower $\sigma_f$ values than the rest of the stem (Fig. \ref{fig:FTLE_section}c, d), although still higher than the background ($\sigma_f \ge$ 0.01 s$^{-1}$). However, the source region of the center of the stem of the plume is at the very base of the bottom TBL, extending laterally far away from the plume \cite{Cagney2015}. It should correspond to a large separation between a pair of tracers and ideally lead to high $\sigma_f$. The discrepancy between expected high values and the computed lower ones, occurs because the bottom TBL is largely removed from the SSPIV domain. The backward-in-time numerical tracers along the stem center will be advected downward and out of the SSPIV domain. When they exit, their last positions, i.e., the base of the plume, will be used for the FTLE calculation here, which leads to a smaller than expected separation and a lower $\sigma_f$. The tracers have not yet separated too far from each other laterally as they would if we were capturing them at  the base of the bottom TBL. The near-zero $\sigma_f$ seen at the base of the plumes (Fig. \ref{fig:FTLE_section}c, d) is a similar artifact due to tracers leaving the domain. More velocity interpolation steps can help reduce these artifacts. Overall, the limited resolution and temporal interpolation is the dominant error source for FTLE, with complex effects on the flow map, as well as the stretching term $\mathrm{max} \left( \frac{||\delta \boldsymbol{x}(t)||}{||\delta \boldsymbol{x}(t_0)||} \right)$. 

\begin{figure}
  \centerline{\includegraphics[width=\linewidth]{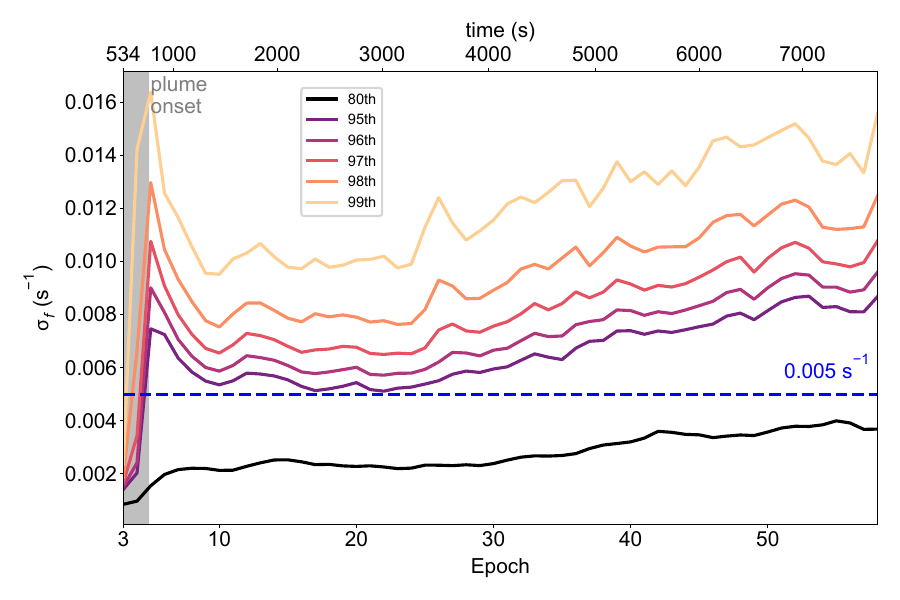}}
  \caption{Per-epoch percentile distribution of the $\sigma_f$ field across the SSPIV domain over time. The 80th percentile (black line) is a representative percentile of the background stretching and remains stably below $0.004 s^{-1}$ across all epochs. The 95th–99th percentiles form the plume FTLE ridge band. The threshold $\sigma_f = 0.005 s^{-1}$ (dashed blue) sits in the gap between the background and the plume band after the onset of the first batch of plumes; the gap remains above 0.002 $s^{-1}$ or more over time. The threshold value is derived as the minimum of the 95th percentile across post-onset epochs. The pre-onset region is shaded. Top x axis: dimensional time in seconds.}
\label{fig:FTLE_threshold}
\end{figure}

We also present the position of the $\sigma_f$ threshold in the FTLE percentile space over time in Fig. \ref{fig:FTLE_threshold}. The threshold $\sigma_f = 0.005 s^{-1}$ is the minimum value for the 95th percentile after the onset of the first batch of plumes. It is clearly separated from the background (represented by the 80th percentile) and the plume FTLE ridges (represented by the 95-99th percentiles). Note that the first batch of plumes just initiated at the end of Epoch 3. They reach less than 1/3 of the tank height at the end of epoch 4. The $\sigma_f$ was calculated by the stretching of the fluid parcel, divided by the time elapsed from the beginning, including the quiescent stage (before the end of Epoch 3). For these times the $\sigma_f$ percentiles are considerably smaller than the rest of the experiment, especially Epoch 5, which gives the largest percentiles. We therefore consider the plume onset as Epoch 5 for the purpose of $\sigma_f$ threshold discussion here.

\bibliography{agusample.bib}

%
%
%
%
%

\end{document}